\documentclass[aps,prl,floatfix,twocolumn,reprint,amsmath,amssymb,superscriptaddress]{revtex4-1}
\usepackage{amsfonts}
\usepackage{mathrsfs}
\usepackage{amsmath}
\usepackage{color}
\usepackage{natbib}
\usepackage{graphicx}
\usepackage{bm}
\usepackage{amssymb}
\usepackage{xspace}
\usepackage{epstopdf}
\usepackage{dcolumn}
\usepackage{longtable}
\usepackage{multirow}
\usepackage[colorlinks=true, letterpaper=true, pdfstartview=FitV, linkcolor=blue, citecolor=blue, urlcolor=blue]{hyperref}
\UseRawInputEncoding

\makeatletter

\newcommand{\Rmnum}[1]{\expandafter\@slowromancap\romannumeral #1@}
\makeatother

\begin{document}

\title{Two-dimensional double-kagome-lattice nitrogene: a direct band gap semiconductor with nontrivial corner state}
\date{\today}

\author{Wenzhang Li}
\affiliation{College of Physics and Electronic Engineering, Center for Computational Sciences, Sichuan Normal University, Chengdu, 610068, China}

\author{Qin He}
\affiliation{College of Physics and Electronic Engineering, Center for Computational Sciences, Sichuan Normal University, Chengdu, 610068, China}

\author{Xiao-Ping Li}
\affiliation{School of Physical Science and Technology, Inner Mongolia University, Hohhot 010021, China}

\author{Da-Shuai Ma}
\affiliation{Institute for Structure and Function $\&$ Department of Physics, Chongqing University, Chongqing 400044, China}

\author{Botao Fu}
\email[]{fubotao2008@gmail.com}
\affiliation{College of Physics and Electronic Engineering, Center for Computational Sciences, Sichuan Normal University, Chengdu, 610068, China}

\begin{abstract}
Based on first-principles calculations, we predict that nitrogen atoms can assemble into a single-layer double kagome lattice (DKL), which possesses the characteristics of an intrinsic direct band gap semiconductor, boasting a substantial band gap of 3.460 eV.
The DKL structure results in a flat valence band with high effective mass and a conduction band with small effective mass comes from Dirac electrons. These distinctive band edges lead to a significant disparity in carrier mobilities, with electron mobility being four orders of magnitude higher than that of holes. The presence of flat band in DKL-nitrogene can be further discerned through the enhanced optical absorption and correlated effects as exemplified by hole-induced ferromagnetism. Interestingly, DKL-nitrogene exhibits inherent second-order topological states, confirmed by a non-trivial second Stiefel-Whitney number and the presence of 1D floating edge states and 0D corner states within the bulk band gap.
Additionally, the robust N-N bonds and the lattice's bending structure ensure thermodynamic stability and mechanical stiffness. These attributes make it exceptionally stable for potential applications in nano-devices.

\end{abstract}

\maketitle

\section{Introduction}
In recent years, the exploration of two-dimensional (2D) element (Xenes) beyond graphene has sparked a revolutionary shift in materials science and condensed matter physics\cite{bhimanapati2015recent,molle2017buckled,xie2021chemistry,mannix2017synthesis}. These atomically thin substances exhibit exceptional properties, offering a wealth of opportunities for novel electronic and optoelectronic devices.
Among these 2D materials, the group-VA Xenes\cite{zhang2018recent,khan2021novel,xia2019black} (nitrogene, phosphorene\cite{li2014black,liu2014phosphorene,bluePPhysRevLett.112.176802}, arsenene\cite{mardanya2016four,zhong2018thickness,kamal2015arsenene,kamal2015arsenene}, antimonene\cite{shao2018epitaxial,wu2017epitaxial,wang2015atomically}, and bismuthene\cite{lu2014topological,liu2011stable,drozdov2014one}) that crystal in buckled hexagonal or puckered rectangular lattice have garnered considerable attention due to their intrinsic semiconducting band gap\cite{zhang2016semiconducting}, high carrier mobility\cite{qiao2014high} and strongly anisotropic properties\cite{xia2014rediscovering}. Despite the considerable progress in understanding 2D materials, research concerning elemental compounds containing nitrogen remains remarkably scarce\cite{researchnitrogen}. This scarcity is surprising given nitrogene critical importance in various chemical and biological processes. The dearth of studies exploring nitrogen-based single-element materials has hindered the comprehensive understanding of their potential and intriguing properties.

Structure determines property, among diverse 2D crystal structures, the kagome lattice holds particular signifcance.
Kagome-formatted materials are characterized by a lattice arrangement resembling a woven basket, possessing fascinating flat-band and Dirac electron that set them apart from other 2D crystals\cite{PhysRevLett.121.096401,flatdirac,kang2020dirac}. This unique electronic structure gives rise to a range of interesting phenomena such as topological quantum state and many-body physics\cite{yin2018giant,manybodyeffect,yin2022topological}, making them an enticing avenue for further research.
Motivated by the lack of research on nitrogen-based single-element materials and the alluring properties of kagome-structured materials, we endeavor to introduce a novel nitrogen-based monolayer with a kagome-related lattice and explore the realm of this fascinating material and unravel its extraordinary electronic and topological characteristics.

In this work, we theoretically propose a kind of monolayer nitrogen that crystals in a double kagome lattice (DKL), which we term DKL-nitrogene. Notably, this material features Dirac electron in the conduction band and a flat band in the valence band.
This giant electron-hole asymmetric in the band structure results in is a remarkable diversity in carrier mobilities for DKL-nitrogene.
The impact from the flat-band is further revealed in DKL-nitrogene that it not only enhances optical absorption in the ultraviolet region but also induces half-metal ferromagnetism under proper hole doping. Furthermore, DKL-nitrogene is confirmed to be a second-order topological insulator, hosting non-zero second Stiefel-Whitney number, 1D floating edge states, and 0D corner states. Thus, DKL-nitrogene holds great promise for applications in electronics and optoelectronics with the potential to unlock exotic quantum phenomena and inspire future research in the field of 2D single-element materials.


\section{Calculation method}
The first-principles calculations are performed based on density functional theory (DFT) and implemented in Vienna ab initio simulation package (VASP)\cite{PhysRevB.47.558,PhysRevB.49.14251,kresse1996efficiency,PhysRevB.54.11169} with projector-augmented-wave (PAW)\cite{PhysRevB.50.17953} and Perdewe-Burke-Ernzerhof (PBE)\cite{PhysRevLett.77.3865} exchange correlation potential. Additionally, the HSE06 hybrid functionals\cite{krukau2006influence} is considered for better description of the band gap and optical absorption. A cut-off energy of 500 eV and a $k$-mesh of $15\times 15 \times 1$ are selected\cite{PhysRevB.13.5188}. The convergence criteria for total energy and force are set at  \begin{math}10^{\small -6}\end{math} eV and -0.001 eV/{\AA}, respectively. The phonon spectrum is calculated using the PHONOPY package\cite{togo2015first}. The tight-binding (TB) Hamiltonian was constructed via MagneticTB code\cite{zhang2023magnetickp}, and the topological edge states were calculated based on Wanniertool code\cite{wanniertool2017}.

\section{Atomic structure and stability of DKL-nitrogene}
\begin{figure}
\centering
\includegraphics[width=0.5\textwidth]{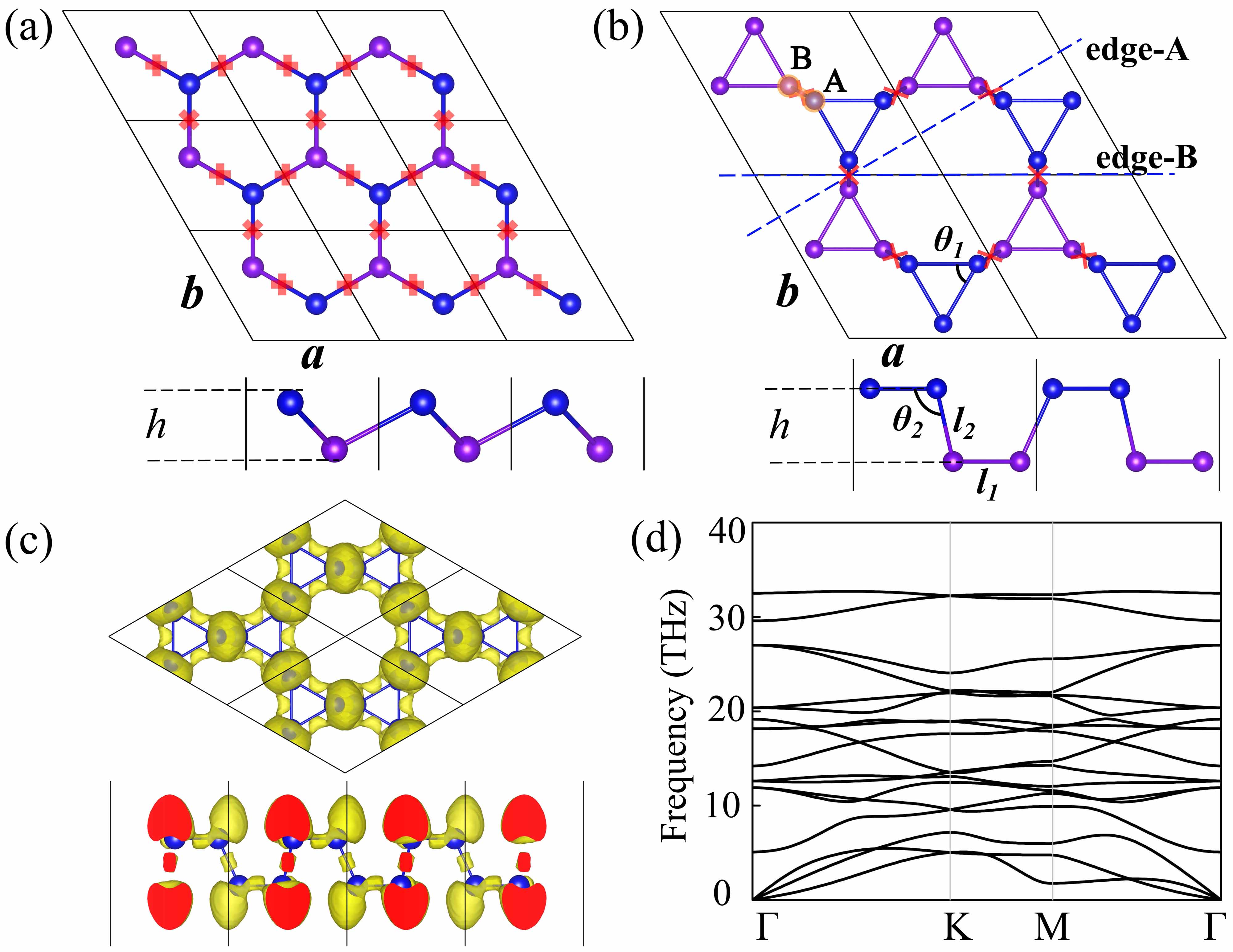}
\caption{ (a) Top and side views of $hex$-nitrogene. (b) Top and side views of DKL-nitrogene. The A and B sublattices are located at different heights, distinguished by blue and purple colors, respectively. The red crosses at the bonding centers between two N-atoms form a typical kagome lattice. (c) Electron local function diagram (ELF) of DKL-nitrogene. (d) Phonon dispersion of DKL-nitrogene.}\label{FIG1}
\end{figure}

In contrast to other pnictogens, nitrogen typically remains in its gaseous state and only forms the nonmolecular solid phase under extreme conditions. Recently, a monolayer nitrogen with buckled honeycomb structure\cite{hex-nitrogene} was predicted stable at room temperature, referred as ``nitrogene". As depicted in Fig.~\ref{FIG1}(a), this hexagonal nitrogene features three-coordinated A/B sublattices with a buckling height ($h$) of 0.70 {\AA}, resembling blue phosphorene. Theoretically, a standard kagome lattice represents the line graph of the honeycomb lattice\cite{dklPhysRevB.99.100404}, which can be achieved by placing an atom (red crossing in Fig.~\ref{FIG1}(a)) at the midpoint of each side of the regular hexagon. However, the arrangement of atoms in the kagome lattice violates the typically triple-coordination bonding rule of pnictogens, resulting in structural instability. To restore the triple-coordination bonding, each atom can be substituted with a binary component (A@B), leading to the formation of a double kagome lattice (DKL)\cite{DKLaelm.202300212} or  diatomic kagome lattice\cite{DKLPhysRevMaterials.5.084203}, as illustrated in Fig.~\ref{FIG1}(b). This DKL can also be directly obtained from the honeycomb lattice by simply replacing every N-atom with a N-trimer. Interestingly, such DKL structures have been discovered in several covalent organic frameworks\cite{organicgrid} but have rarely been discussed in elemental materials\cite{dklPhysRevB.98.035135,bftC9TC06132K}.

In the optimized DKL-nitrogene, each N-atom remains representative of a three-coordination of group-VA elements with two different bonding lengths ($l_1$,$l_2$) and bonding angles (${\theta}_1$, ${\theta}_2$) as shown in Fig.~\ref{FIG1}(b). The optimized lattice constant of DKL-nitrogene is about 4.04 {\AA}, with intra-trimer and inter-trimer bonding length of 1.48 {\AA} and 1.53 {\AA}, respectively, which is similar to that of $hex$-nitrogene\cite{hex-nitrogene}. The intra-trimer angle ${\theta}_1$ is constrained to $60^{\circ}$ and the inter-trimer angle ${\theta}_2$ is about $111^{\circ}$. The buckling height between top and bottom N-atoms is about 1.40 {\AA}, significantly larger than that of $hex$-nitrogene (0.70 {\AA}), due to strong $sp^3$ hybridization. It's worthwhile that the crystal structure of DKL-nitrogene also bears a resemblance to that of black phosphorene when viewed from the side, sharing a similar puckered geometry.

By visualizing the electron local function (ELF) in Fig.~\ref{FIG1}(c), a robust covalent bonding nature between N-N atoms becomes evident, and the presence of a lone pair electron on the N-atom is distinctly illustrated. The dynamic stability of DKL-nitrogene is unequivocally affirmed by the phonon spectrum illustrated in Fig.~\ref{FIG1}(d), wherein no imaginary frequencies are observed throughout the entire Brillouin zone. Additionally, the thermal and mechanical stability of DKL-nitrogene are firmly established through ab initio molecular dynamics simulations in Fig. S1 and adhering to Born's mechanical stability criteria\cite{mechanicalstability} as elaborated in Tab. S1, and corresponding Young's modulus and Poisson's ratio are shown in Fig. S3 in the supplementary information.
\section{Electronic structure of DKL-nitrogene}
\begin{figure*}[t]
\includegraphics[width=16.5 cm]{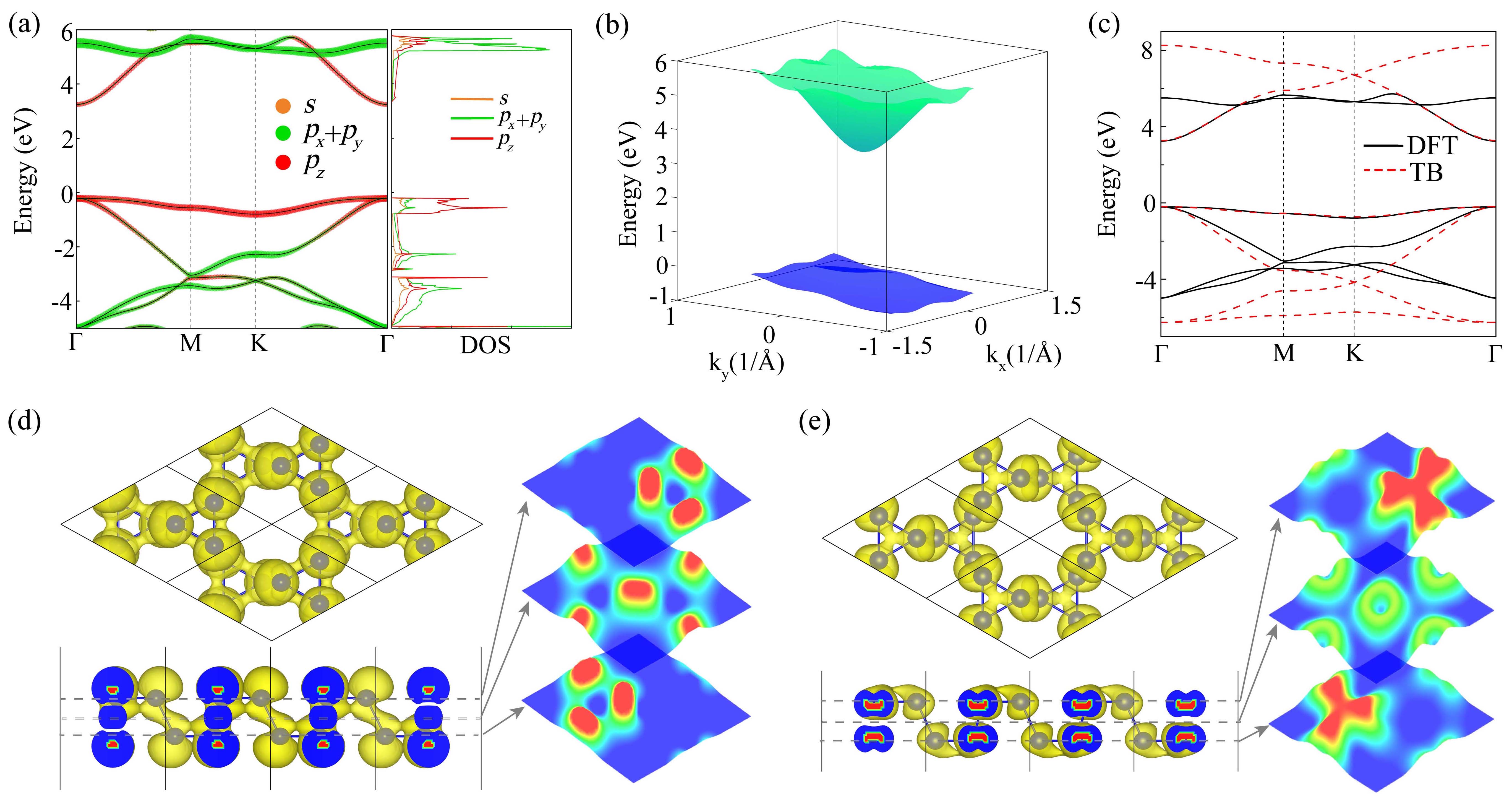}
\caption{(a) Orbital-resolved band structure and density of states (DOS) of DKL-nitrogene. (b) 3D view of the HVB and LCB of DKL-nitrogene. (c) Band structure from TB model with parameters: $e_0=-0.32$, $t_1=3.0$, $t_2=2.86$, $t_5=-0.18$. Here, $t_n$ denotes the $n$th order hopping parameter. (d) Charge density of the HVB. (e) Charge density of the LCB.
}\label{FIG2}
\end{figure*}
The electronic band structure of DKL-nitrogene is determined through calculations incorporating the HSE06 correction, as depicted in Fig.~\ref{FIG2}(a). Unlike its hexagonal allotrope, which is recognized as an indirect insulator (5.900 eV, HSE06)\cite{hex-nitrogene}, DKL-nitrogene exhibits a distinct behavior as a promising direct band gap semiconductor, with both the valence band maximum (VBM) and conduction band minimum (CBM) at the $\Gamma$  point. The calculated band gap is approximately 3.460 eV at the HSE06 level. Due to its substantial direct band gap, DKL-nitrogen holds potential utility within the visible and ultraviolet spectral ranges. When examining the orbital-projected band structure, it becomes apparent that the highest valence band (HVB) and the lowest conduction band (LCB) are predominantly composed of $p_z$ orbitals, with minor contributions from $p_{x,y}$ and $s$ orbitals. Intriguingly, despite the similar orbital composition, HVB and LCB exhibit strikingly different dispersion behaviors. As clearly illustrated in Fig.~\ref{FIG2}(b) (in a 3D view), the former conforms to a conventional conduction band pattern with typically parabolic dispersion, while the latter shows nearly flat dispersion with a small bandwidth of 0.590 eV.

To further elucidate the potential mechanism underlying this phenomenon, we have plotted the 3D charge density for the HVB and the LCB along with 2D cross-sectional views, as shown in Figs.~\ref{FIG2}(d) and \ref{FIG2}(e), respectively. In the case of the HVB, the three intra-trimer N-atoms form in-plane anti-bonding states, with reduced charge density between them, while two inter-trimer N-atoms form out-of-plane bonding states. Remarkably, these inter-trimer bonding centers, as displayed in the 2D cross-sectional view, actually give rise to a typical kagome lattice pattern, ultimately leading to the emergence of a flat band. Conversely, for the LCB, the three intra-trimer N-atoms engage in in-plane bonding states, resulting in trefoil-like charge patterns, while two inter-trimer N-atoms form out-of-plane anti-bonding states with reduced charge density between them. As a whole, the centers of the two trefoils on the top and bottom planes create a buckled honeycomb lattice. This lattice configuration can account for massive LCB, including the presence of the Dirac cone at the K point.

Because the band edges are mainly formed by $p_z$-orbital, a simple single-orbital toy model can be written as:
\begin{eqnarray}
H=\sum_{i,j} t_{n}c^{\dagger}_{i}c_{j}+h.c.
\end{eqnarray}
, where the $c^{\dagger}_{i}$ and $c_j$ are the creation and annihilation operators at sites $i$ and $j$, respectively. The $t_n$ is the $n$-th order hopping term. In the supplementary information (Fig. S2), we extensively discussed the impact of various hopping parameters on the band structure. By fitting these parameters, we obtained the band structure of DKL-nitrogene, as illustrated in Fig.~\ref{FIG2}(c). Six $p_z$ orbitals on the DKL form two sets of bands\cite{liufengPhysRevLett.130.186401}. One set (conduction band) involves two coupled bands that cross with each other at $K$ point with the energy minimum point at $\Gamma$ point whereby the band is quite dispersive. The other set (valence band) involves four coupled bands that form a hourglass-like structure where two middle bands cross each forming Dirac cone at $K$, which is further sandwiched by two flat bands from the top and bottom sides. The weak dispersion of flat highest valence band is dominated by 5-$th$ order hopping term ($t_5$) in our TB model.

\section{Anisotropic and electronic-hole asymmetric carrier mobility}
The carrier mobility of a material greatly affects its applicability as an electronic device. With approximation of longitudinal acoustic wave, the phonon-related carrier mobility of 2D structure can be evaluated by a general formula\cite{qiao2014high}:
\begin{eqnarray}
\mu_{x,y}=\frac{e\hbar^3 C_{x,y}}{k_B Tm_{x,y}^{*} (m_x^* m_y^* )^{1/2}D_{x,y} ^2 },\label{mobilityf1}
\end{eqnarray}
the $m_{x,y}^{*}$ represents the effective mass of carriers, $C_{x,y}$ stands for the modulus of elasticity, and $D_{x,y}$ is the deformation potential constant. Here the subscript $x$ and $y$ denote the zigzag and armchair directions, respectively, and the temperature is set at 300 K.
Thereinto, the deformation potential is defined as,
\begin{equation}
D_{x,y}=\frac{\partial E_{n\boldsymbol{k}} }{\partial {\varepsilon_{x,y}}},
\end{equation}
where $E_{n\boldsymbol{k}}$ represents the electronic eigenvalue for a band with index $n$ and wave vector $\boldsymbol{k}$ at VBM or CBM, while ${\varepsilon_{x,y}}$ denotes to the uniaxial strain along $x$ or $y$ direction. The deformation potential characterizes the energy shift in response to external strain perturbations, primarily accounting for phonon scattering from breathing modes.
The effective mass $m^{*}$ is obtained by fitting band dispersion with formula:\begin{equation}
\frac{1}{m_{x,y}^{*}}=\frac{1}{\hbar^{2}}\frac{\partial^{2}E_{n\boldsymbol{k}}}{\partial k_{x,y}^{2}}.
\end{equation}
The modulus of elasticity is calculated as,
\begin{equation}
C_{x,y}=\frac{1}{S_{0}}\frac{\partial^{2}E}{\partial \large\varepsilon_{x,y}^{2}},
\end{equation}
where $E$ is the total energy, and $ S_{0} $ is the area of the unit cell.

\begin{table*}
\centering
\begin{tabular}{ c | c | c | c | c | c | c | c | c }
\hline
Carriers & $m_{x}^{*}/m_{0}$ & $m_{y}^{*}/m_{0}$ & $D_{x}$(eV) & $D_{y}$(eV) & $C_{x}$(Jm\begin{math}^{\small -2}\end{math}) & $C_{y}$(Jm\begin{math}^{\small -2}\end{math}) & $\mu_{x}$(cm\begin{math}^{\small 2}\end{math}V\begin{math}^{\small -1}\end{math}s\begin{math}^{\small -1}\end{math}) & $\mu_{y}$(cm\begin{math}^{\small 2}\end{math}V\begin{math}^{\small -1}\end{math}s\begin{math}^{\small -1}\end{math}) \\
\hline
Electron & 0.34 & 0.34 & 1.30 & 1.42 & 95.15 & 94.78 & 10390.00 & 8674.21 \\
\hline
Hole & 15.30 & 16.90 & 2.96 & 2.93 & 95.15 & 94.78 & 0.94 & 0.87 \\
\hline
\end{tabular}
\caption{Effective mass $m^*_{x,y}$, deformation potential constant $D_{x,y}$, elastic modulus $C_{x,y}$ and carrier mobility $\mu_{x,y}$ of DKL-nitrogene, where $x$ and $y$ represent the transport direction. The $m_0$ is the free-electron mass.} \label{mobility}
\end{table*}
Based on the calculated effective mass ($m^{*}$), elastic modulus constants ($C_{x,y}$), and deformation potential constant ($D_{x,y}$) for the CBM and VBM at $\Gamma$ point, we have determined the mobilities of both electrons and holes, as presented in Table.~\ref{mobility}. One of the most striking features of the mobility is the substantial asymmetry between electrons and holes, with electron mobility being four orders of magnitude greater than hole mobility. For instance, electron mobility ($\mu_x$/$\mu_y$) reaches a substantial value of 1.04$\times$10$^4$/0.87$\times$10$^4$ cm$^2$V$^{-1}$s$^{-1}$, while hole mobility is only 0.94/0.87 cm$^2$V$^{-1}$s$^{-1}$.
We attribute this asymmetry in carrier mobility primarily to the intrinsic electron-hole asymmetry within the unique band structure. This band structure displays a dispersive conduction band with a Dirac cone, resulting in a small effective mass ($m_{x}^{*}$/$m_{y}^{*}$) for electrons (0.34/0.34$m_0$). In contrast, the valence band is exceedingly flat, leading to a large effective mass for holes (15.30/16.90$m_0$), as listed in Table.~\ref{mobility}. Furthermore, aside from the difference in effective mass, the deformation potential ($D_x/D_y$) of electrons is only 1.30/1.42 eV, notably smaller than that of holes (2.96/2.93 eV). Therefore, the significant contrast in effective mass and in deformation potential between electrons and holes ultimately result in a substantial disparity in carrier mobility, as clearly indicated by Eq.~(\ref{mobilityf1}). This pronounced difference in carrier mobility has the potential to enhance the efficiency of separating photo-generated electrons and holes\cite{li2013spatial}, opening up broad applications in electronic devices, optoelectronics, and photocatalysis.

\section{Enhanced optical absorption of DKL-nitrogene}

Based on the linear response theory and independent particle approximation\cite{linearresponsetheory}, we calculate the frequency-dependent dielectric function ${\varepsilon (w)}={\varepsilon_1 (w)}+i{\varepsilon_2 (w)}$. The ${\varepsilon_2(w)}$ represents the imaginary part of the dielectric function that can be derived from the inter-band optical transition, and the ${\varepsilon_1(w)}$ is the real part of dielectric function that can be derived from ${\varepsilon_2(w)}$ according to Kramers-Kronig relationship\cite{PhysRevB.73.045112}.
To avoid the influence from the vacuum layer, it's more reasonable to calculate the 2D optical absorbance $A(w)$ from corresponding optical conductivity $\sigma_{2D}(w)$\cite{matthes2014optical,PhysRevB.94.205408}.
Thereinto, the 2D optical conductivity is derived from the corresponding 3D counterpart through the equation,
 \begin{equation}
\sigma_{2D}(w)=L\sigma_{3D}(w),
\end{equation}
where $L$ represents the total length of unitcell along out-plane direction, and the $\sigma_{3D}(w)$ is derived from DFT obtained dielectric function of 2D system with sufficient evacuum layer as
\begin{equation}
\sigma_{3D}(w)=i[1-\varepsilon(w)]\varepsilon_{0}w,
\end{equation}
where $ \varepsilon_{0} $ is the permittivity of vacuum and $ w $ is the frequency of the incident wave.
Then, the normalized absorbance $ A(w) $ of a two-dimensional material at normal incidence can be expressed as
\begin{equation}
A(w)=\frac{Re\widetilde{\sigma}}{|1+\widetilde{\sigma}/2|^{2}} ,
\end{equation}
where $ \widetilde{\sigma}=\sigma_{2D}(w)/\varepsilon_{0}c $ is the normalized 2D optical conductivity ($c$ is the speed of light).

To elucidate the impact of the unique electronic structure on the optical absorption coefficient, we conducted comparative calculations for all predicted monolayer nitrogenes, including DKL-nitrogene, $hex$-nitrogene\cite{hex-nitrogene}, $oct$-nitrogene\cite{oct-nitrogene} and ZS-nitrogene\cite{ZS-nitrogene}, whose crystal structure and band structure are given in Fig. S4 in supplementary information. Remarkably, DKL-nitrogene demonstrates significantly enhanced optical absorption in the range from 200 nm to 600 nm. In particular, the DKL-nitrogene exhibits much larger optical absorption capabilities in the near-ultraviolet range (200-370 nm), when compared to nitrogen allotropes.
Only ZS-nitrogene displays larger optical absorption within the 370-600 nm range due to its smaller band gap (2.451 eV).
Furthermore, considering the diverse band gaps and band edge characteristics of these four materials, we specifically compared their optical absorption near the band edge, as illustrated in the inset of Fig.~\ref{FIG3}(a). Within 35 nm range, we observed that DKL-nitrogene exhibits the highest absorption coefficient at the band edge position compared to the other three structures. This phenomenon can be attributed to its unique band edge electronic structures. Specifically, in DKL-nitrogene, the flat valence band provides a high electron density, contributing to substantial light absorption at the band edge.


\section{Flat-band ferromagnetism induced by hole doping in DKL-nitrogene}
\begin{figure}
\centering
\includegraphics[width=0.5\textwidth]{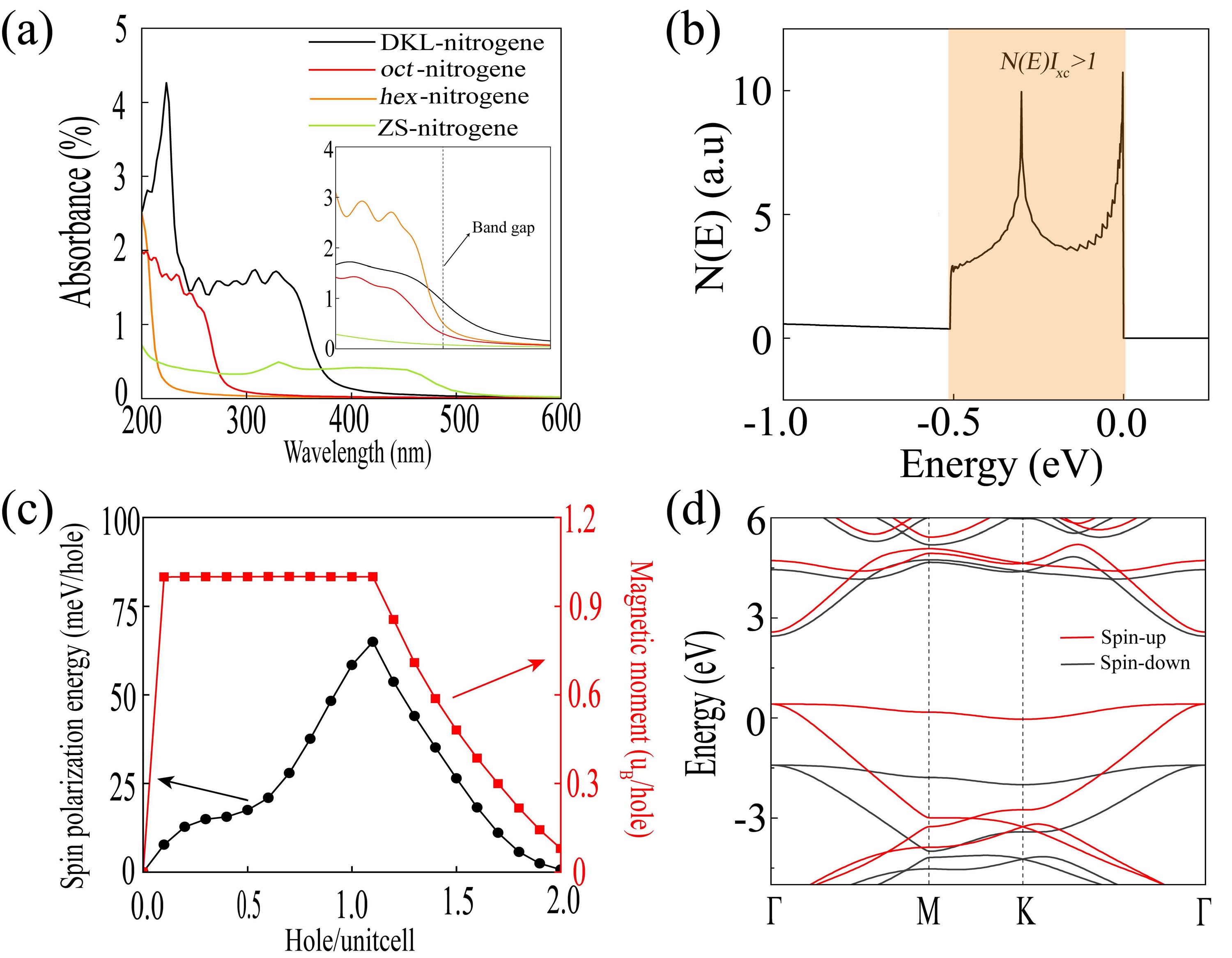}
\caption{(a) The optical absorption coefficient for DKL-nitrogene, $hex$-nitrogene, $oct$-nitrogene and ZS-nitrogene. Inset is the absorption coefficient around band gap. (b) Total DOS of DKL-nitrogene around Fermi level. (c) Spin polarization energy and spin magnetic moment versus hole doping concentration ($n$) for DKL-nitrogene. (d) Spin-polarized band structures under $n$=1.2 per unitcell.}\label{FIG3}
\end{figure}

The kagome lattice stands out due to its uniquely flat band structure, which offers a platform for exploring intriguing many-body phenomena.
For instance, the valence band observed in DKL-nitrogene contributes to an extremely large DOS just below Fermi level as depicted in Fig.~\ref{FIG3}(b).
By manipulating hole concentrations within this context, it is possible to potentially induce electronic instabilities. These instabilities, in turn, could lead to various outcomes, such as magnetic phase transitions or Wigner crystallization\cite{dklPhysRevB.98.035135}. To further verify this concept, our study delves into the behavior of the spin magnetic moment and spin polarization energy as a function of increasing hole density per unit cell ($n$), as presented in Fig.~\ref{FIG3}(c).
Remarkably, the spin magnetic moment emerges even at exceedingly low doping densities and quickly reaches a saturation value of 1.0 $\mu_B$. This value remains constant up to $n$=1.2, after which it gradually diminishes to zero. The spin polarization energy, defined as the energy difference between the ferromagnetic (FM) and non-magnetic (NM) states, exhibits a prominent positive peak across the entire range of doping levels. This peak strongly suggests a pronounced energy preference for the ferromagnetic state.

The emergence of itinerant magnetism induced by doping can be effectively understood through the lens of the Stoner criterion\cite{Stonercriterion}:
\begin{equation}
N(E_\text{F})I_{xc}>1,
\end{equation}
where $N(E_\text{F})$ represents the DOS at the Fermi energy in the non-magnetic (NM) state, and $I_{xc}$ is the Stoner parameter. The Stoner parameter can be approximated using the relationship\cite{stonerPhysRevB.78.104402} $I_{xc}=\Delta_{xc}/\mathbf{M}$, where $\Delta_{xc}$ signifies the energy of exchange splitting between the two spin channels, and $\mathbf{M}$ denotes the total magnetic moment.
In the case of DKL-nitrogene, $I_{xc}\simeq 0.65$. Evidently, the condition $N(E_\text{F})I_{xc}>1$ is satisfied across the entire filling range of the flat band, as illustrated in Fig.~\ref{FIG3}(b).
Notably, the saturation magnetic moment of 1 ¦ÌB per hole indicates a 100$\%$ spin polarization, confirming the presence of half-metallic states. For instance, when considering $n$=1.2, this state exhibits the most substantial spin polarization energy, quantified at 65 meV per hole. As illustrated in Fig.~\ref{FIG3}(d), it becomes evident that only the spin-up band crossing the Fermi level leads to the formation of a half-metallic ferromagnetic (FM) state.

\section{Second-order topological insulator with edge and corner states}
In the realm of 2D materials, a diverse range of compounds boasting substantial band gaps are systematically categorized into second-order topological insulators (SOTIs)\cite{SOTIWladimir2017,PhysRevB.96.245115}. Setting them apart from conventional 2D topological insulators that typically exhibit 1D conductive edge states bridging the conduction and valence bands, these 2D SOTIs require no SOC and uniquely harbor 1D gapped edge state as well as 0D corner states.
These remarkable corner states emerge specifically when these materials are engineered into quantum dots with precisely tailored geometries\cite{PhysRevLett.123.256402,SOTIgraphyneNL,SOTIPhysRevB.104.245427}.

\begin{figure}
\centering
\includegraphics[width=0.5\textwidth]{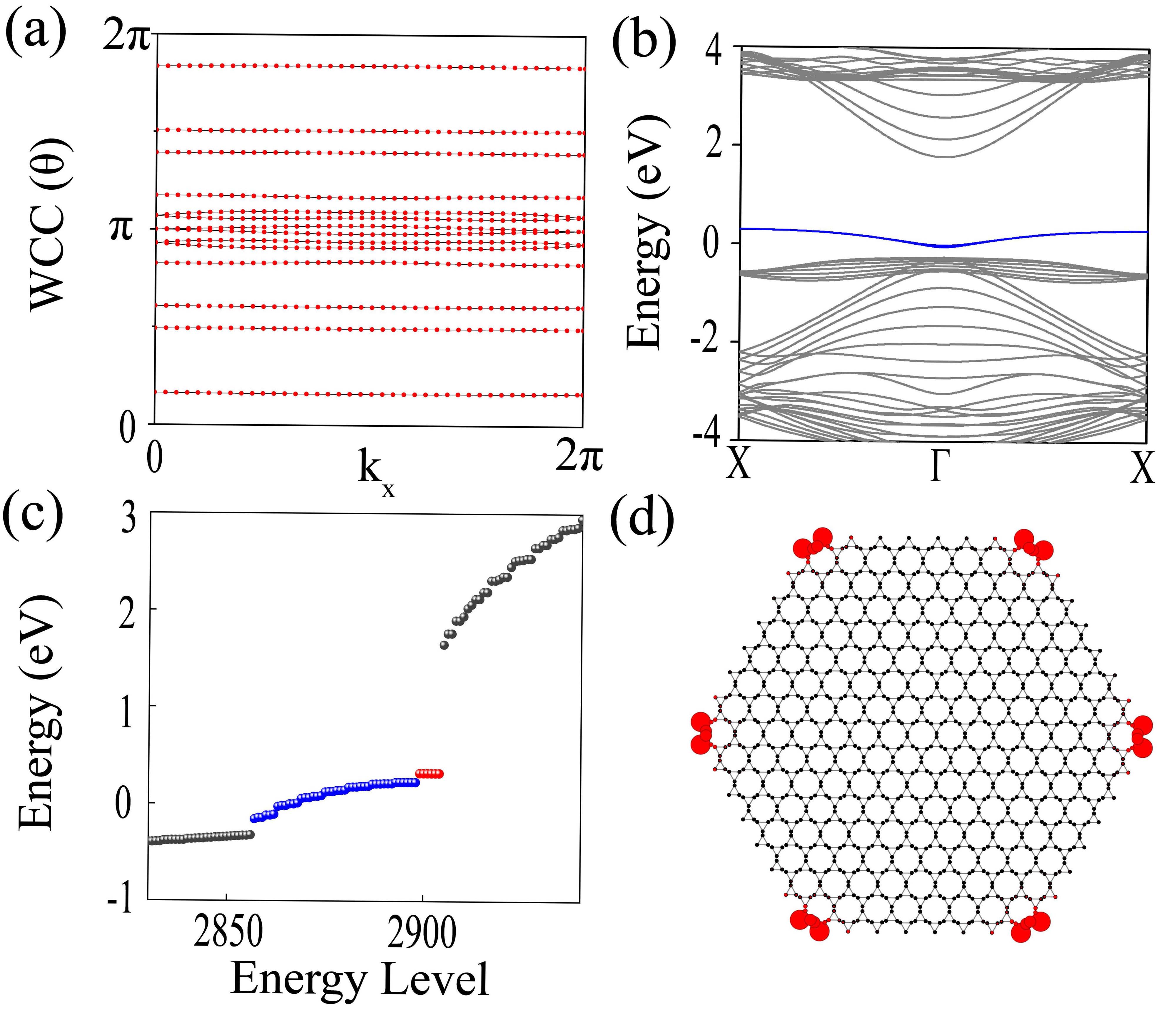}
\caption{(a) The Wilson loop of DKL-nitrogene. The number of crossing on $\theta$=$\pi$ is 1, therefore the $w_2$ is 1. (b) The  energy spectrum of 1D nanoribbon with type-A edge configuration. The mid gap edge state is marked by blue color. (c) The energy spectrum of 0D hexagonal flake (N=1152) composed of type-A edges. The grey, blue and red dots stand for the bulk, edge and corner states, respectively. (d) The charge spatial distribution of corner states.}\label{FIG4}
\end{figure}

To characterize SOTIs, a novel topological invariant known as the second Stiefel-Whitney number ($w_2$) has been introduced\cite{Ahn2019,PhysRevLett.118.056401}. For the DKL-nitrogene with inversion symmetry, $w_2$ can be calculated directly using parity eigenvalues as follows:
\begin{eqnarray}
(-1)^{w_2}=\prod_{i=1}^{4}(-1)^{{\lfloor} N_{occ}^{-}(\Gamma_i)/2{\rfloor}},
\end{eqnarray}
where $N_{occ}^{-}$ denotes the number of occupied bands with odd parity at the time reversal momentum ($\Gamma_i$), and ${\lfloor}  {\rfloor}$ is the floor function. The calculated $w_2$=1 for DKL-nitrogene indicates that it is a 2D SOTI.
Moreover, this non-zero second Stiefel-Whitney number is also confirmed by the nested Wilson loop spectrum\cite{wieder2018axion}, as shown in Fig.~\ref{FIG4}(a), where the spectrum exhibits a crossing point at $k_x=0$, $\theta=\pi$, indicating $w_2=1$.

To unveil the non-trivial edge states and corner states, two types of boundaries (type-A and type-B) with different atomic geometries are chosen, as shown in Fig.~\ref{FIG1}(a). Firstly, for the 1D ribbon along the type-A direction, only one branch of floating edge state exists in the middle of the bulk band gap, as displayed in Fig.~\ref{FIG4}(b). Next, we explore the 0D corner state for a finite-sized hexagonal flake (N=1152) containing six type-A edge configuration. The discrete energy spectrum of DKL-nitrogene flake is shown in Fig.~\ref{FIG4}(c), with bulk, edge, and corner states differentiated by the gray, blue, and red colors, respectively. Above the edge states, there are six degenerate corner states related by the $C_6$ symmetry, with their charge being equally localized on the six corners of the hexagonal flake, as clearly shown in Fig.~\ref{FIG4}(d). Therefore, considering doping this flake with integer electrons to these 6-fold-degenerate corner states\cite{SOTIwangzf2022}, each corner will hold a fractionally quantized corner charge of $\frac{e}{6}$.
On the other hand, if we construct a hexagonal flake composed of uniform type-B edges (N=1098), there exist three groups of six-degenerate corner states (see Fig. S8 in the supplementary information). One group sits at zero energy between the edge states, while the other two groups sit above and below the edge states.

\section{Discussion and conclusion}
The distinctive structure of 2D materials enables us to finely manipulate their electronic structures using external electric, light, and strain fields. As a result, we have delved deeper into the manipulation of the band structure and light absorption of DKL-nitrogene through strain engineering in Figs. S5-S7 in supplementary information. We initially unveiled that DKL-nitrogene can endure a maximum uniaxial strain of 10$\%$, a threshold comparable to that of other two-dimensional materials in Fig. S5. Moreover, it exhibits an out-of-plane negative Poisson's ratio, reminiscent of black phosphorene's behavior\cite{BPPoissonratio}. More importantly, the value of band gap can be significantly controlled by strain while maintaining its intrinsic direct band-gap characteristics in Fig. S6. Through incremental tensile strain, we observed the band gap decreasing from 3.460 eV to 2.310 eV. Concurrently, the peak absorption wavelength shifted from the near-ultraviolet range to the visible light spectrum in Fig. S7. These exceptional and adjustable optical characteristics form the basis for potential applications of DKL-nitrogene in the field of optoelectronic devices.

In conclusion, we present a stable nitrogen-based monolayer structure characterized by a unique double kagome lattice configuration, offering a platform with intriguing properties. We unveil the asymmetric nature of the parabolically dispersing conduction band and the nearly flat valence band is revealed. Furthermore, we explore the implications of this flat-band structure, emphasizing its role in enhancing optical absorption and facilitating the emergence of flat-band-induced ferromagnetism. Beyond this, we also uncover the underlying higher-order topology present in DKL-nitrogene by calculating the second Stiefel-Whitney number and demonstrating the existence of edge and corner states.
With its exceptional mechanical, optical, and topological attributes, the DKL-nitrogene may pave the way for possible applications in various fields, particularly nano-device technology.

\begin{acknowledgements}
This work is supported by the National Natural Science Foundation of China (NSFC, Grants No. 12304086, No. 12204330, No. 12204074), and Dr. B. Fu also the Sichuan Normal University for financial support (Grant No. 341829001). The numerical computations were performed at the Hefei advanced
computing center, and this research was also supported by the High Performance Computing Center of Sichuan Normal University.

\end{acknowledgements}

%


\begin{thebibliography}{68}%
\makeatletter
\providecommand \@ifxundefined [1]{%
 \@ifx{#1\undefined}
}%
\providecommand \@ifnum [1]{%
 \ifnum #1\expandafter \@firstoftwo
 \else \expandafter \@secondoftwo
 \fi
}%
\providecommand \@ifx [1]{%
 \ifx #1\expandafter \@firstoftwo
 \else \expandafter \@secondoftwo
 \fi
}%
\providecommand \natexlab [1]{#1}%
\providecommand \enquote  [1]{``#1''}%
\providecommand \bibnamefont  [1]{#1}%
\providecommand \bibfnamefont [1]{#1}%
\providecommand \citenamefont [1]{#1}%
\providecommand \href@noop [0]{\@secondoftwo}%
\providecommand \href [0]{\begingroup \@sanitize@url \@href}%
\providecommand \@href[1]{\@@startlink{#1}\@@href}%
\providecommand \@@href[1]{\endgroup#1\@@endlink}%
\providecommand \@sanitize@url [0]{\catcode `\\12\catcode `\$12\catcode
  `\&12\catcode `\#12\catcode `\^12\catcode `\_12\catcode `\%12\relax}%
\providecommand \@@startlink[1]{}%
\providecommand \@@endlink[0]{}%
\providecommand \url  [0]{\begingroup\@sanitize@url \@url }%
\providecommand \@url [1]{\endgroup\@href {#1}{\urlprefix }}%
\providecommand \urlprefix  [0]{URL }%
\providecommand \Eprint [0]{\href }%
\providecommand \doibase [0]{http://dx.doi.org/}%
\providecommand \selectlanguage [0]{\@gobble}%
\providecommand \bibinfo  [0]{\@secondoftwo}%
\providecommand \bibfield  [0]{\@secondoftwo}%
\providecommand \translation [1]{[#1]}%
\providecommand \BibitemOpen [0]{}%
\providecommand \bibitemStop [0]{}%
\providecommand \bibitemNoStop [0]{.\EOS\space}%
\providecommand \EOS [0]{\spacefactor3000\relax}%
\providecommand \BibitemShut  [1]{\csname bibitem#1\endcsname}%
\let\auto@bib@innerbib\@empty
\bibitem [{\citenamefont {Bhimanapati}\ \emph {et~al.}(2015)\citenamefont
  {Bhimanapati}, \citenamefont {Lin}, \citenamefont {Meunier}, \citenamefont
  {Jung}, \citenamefont {Cha}, \citenamefont {Das}, \citenamefont {Xiao},
  \citenamefont {Son}, \citenamefont {Strano}, \citenamefont {Cooper} \emph
  {et~al.}}]{bhimanapati2015recent}%
  \BibitemOpen
  \bibfield  {author} {\bibinfo {author} {\bibfnamefont {G.~R.}\ \bibnamefont
  {Bhimanapati}}, \bibinfo {author} {\bibfnamefont {Z.}~\bibnamefont {Lin}},
  \bibinfo {author} {\bibfnamefont {V.}~\bibnamefont {Meunier}}, \bibinfo
  {author} {\bibfnamefont {Y.}~\bibnamefont {Jung}}, \bibinfo {author}
  {\bibfnamefont {J.}~\bibnamefont {Cha}}, \bibinfo {author} {\bibfnamefont
  {S.}~\bibnamefont {Das}}, \bibinfo {author} {\bibfnamefont {D.}~\bibnamefont
  {Xiao}}, \bibinfo {author} {\bibfnamefont {Y.}~\bibnamefont {Son}}, \bibinfo
  {author} {\bibfnamefont {M.~S.}\ \bibnamefont {Strano}}, \bibinfo {author}
  {\bibfnamefont {V.~R.}\ \bibnamefont {Cooper}},  \emph {et~al.},\ }\href
  {\doibase 10.1021/acsnano.5b05556} {\bibfield  {journal} {\bibinfo  {journal}
  {ACS nano}\ }\textbf {\bibinfo {volume} {9}},\ \bibinfo {pages} {11509}
  (\bibinfo {year} {2015})}\BibitemShut {NoStop}%
\bibitem [{\citenamefont {Molle}\ \emph {et~al.}(2017)\citenamefont {Molle},
  \citenamefont {Goldberger}, \citenamefont {Houssa}, \citenamefont {Xu},
  \citenamefont {Zhang},\ and\ \citenamefont {Akinwande}}]{molle2017buckled}%
  \BibitemOpen
  \bibfield  {author} {\bibinfo {author} {\bibfnamefont {A.}~\bibnamefont
  {Molle}}, \bibinfo {author} {\bibfnamefont {J.}~\bibnamefont {Goldberger}},
  \bibinfo {author} {\bibfnamefont {M.}~\bibnamefont {Houssa}}, \bibinfo
  {author} {\bibfnamefont {Y.}~\bibnamefont {Xu}}, \bibinfo {author}
  {\bibfnamefont {S.-C.}\ \bibnamefont {Zhang}}, \ and\ \bibinfo {author}
  {\bibfnamefont {D.}~\bibnamefont {Akinwande}},\ }\href {\doibase
  10.1038/nmat4802} {\bibfield  {journal} {\bibinfo  {journal} {Nat. Mater.}\
  }\textbf {\bibinfo {volume} {16}},\ \bibinfo {pages} {163} (\bibinfo {year}
  {2017})}\BibitemShut {NoStop}%
\bibitem [{\citenamefont {Xie}\ \emph {et~al.}(2021)\citenamefont {Xie},
  \citenamefont {Zhang}, \citenamefont {Ge}, \citenamefont {Zhu}, \citenamefont
  {Nie}, \citenamefont {Song}, \citenamefont {Lim}, \citenamefont {Zhang},\
  and\ \citenamefont {Prasad}}]{xie2021chemistry}%
  \BibitemOpen
  \bibfield  {author} {\bibinfo {author} {\bibfnamefont {Z.}~\bibnamefont
  {Xie}}, \bibinfo {author} {\bibfnamefont {B.}~\bibnamefont {Zhang}}, \bibinfo
  {author} {\bibfnamefont {Y.}~\bibnamefont {Ge}}, \bibinfo {author}
  {\bibfnamefont {Y.}~\bibnamefont {Zhu}}, \bibinfo {author} {\bibfnamefont
  {G.}~\bibnamefont {Nie}}, \bibinfo {author} {\bibfnamefont {Y.}~\bibnamefont
  {Song}}, \bibinfo {author} {\bibfnamefont {C.-K.}\ \bibnamefont {Lim}},
  \bibinfo {author} {\bibfnamefont {H.}~\bibnamefont {Zhang}}, \ and\ \bibinfo
  {author} {\bibfnamefont {P.~N.}\ \bibnamefont {Prasad}},\ }\href {\doibase
  10.1021/acs.chemrev.1c00165} {\bibfield  {journal} {\bibinfo  {journal}
  {Chem. Rev.}\ }\textbf {\bibinfo {volume} {122}},\ \bibinfo {pages} {1127}
  (\bibinfo {year} {2021})}\BibitemShut {NoStop}%
\bibitem [{\citenamefont {Mannix}\ \emph {et~al.}(2017)\citenamefont {Mannix},
  \citenamefont {Kiraly}, \citenamefont {Hersam},\ and\ \citenamefont
  {Guisinger}}]{mannix2017synthesis}%
  \BibitemOpen
  \bibfield  {author} {\bibinfo {author} {\bibfnamefont {A.~J.}\ \bibnamefont
  {Mannix}}, \bibinfo {author} {\bibfnamefont {B.}~\bibnamefont {Kiraly}},
  \bibinfo {author} {\bibfnamefont {M.~C.}\ \bibnamefont {Hersam}}, \ and\
  \bibinfo {author} {\bibfnamefont {N.~P.}\ \bibnamefont {Guisinger}},\ }\href
  {https://www.nature.com/articles/s41570-016-0014} {\bibfield  {journal}
  {\bibinfo  {journal} {Nat. Rev. Chem.}\ }\textbf {\bibinfo {volume} {1}},\
  \bibinfo {pages} {0014} (\bibinfo {year} {2017})}\BibitemShut {NoStop}%
\bibitem [{\citenamefont {Zhang}\ \emph {et~al.}(2018)\citenamefont {Zhang},
  \citenamefont {Guo}, \citenamefont {Chen}, \citenamefont {Wang},
  \citenamefont {Gao}, \citenamefont {G{\'o}mez-Herrero}, \citenamefont {Ares},
  \citenamefont {Zamora}, \citenamefont {Zhu},\ and\ \citenamefont
  {Zeng}}]{zhang2018recent}%
  \BibitemOpen
  \bibfield  {author} {\bibinfo {author} {\bibfnamefont {S.}~\bibnamefont
  {Zhang}}, \bibinfo {author} {\bibfnamefont {S.}~\bibnamefont {Guo}}, \bibinfo
  {author} {\bibfnamefont {Z.}~\bibnamefont {Chen}}, \bibinfo {author}
  {\bibfnamefont {Y.}~\bibnamefont {Wang}}, \bibinfo {author} {\bibfnamefont
  {H.}~\bibnamefont {Gao}}, \bibinfo {author} {\bibfnamefont {J.}~\bibnamefont
  {G{\'o}mez-Herrero}}, \bibinfo {author} {\bibfnamefont {P.}~\bibnamefont
  {Ares}}, \bibinfo {author} {\bibfnamefont {F.}~\bibnamefont {Zamora}},
  \bibinfo {author} {\bibfnamefont {Z.}~\bibnamefont {Zhu}}, \ and\ \bibinfo
  {author} {\bibfnamefont {H.}~\bibnamefont {Zeng}},\ }\href {\doibase
  10.1039/C7CS00125H} {\bibfield  {journal} {\bibinfo  {journal} {Chem. Soc.
  Rev.}\ }\textbf {\bibinfo {volume} {47}},\ \bibinfo {pages} {982} (\bibinfo
  {year} {2018})}\BibitemShut {NoStop}%
\bibitem [{\citenamefont {Khan}\ \emph {et~al.}(2021)\citenamefont {Khan},
  \citenamefont {Tareen}, \citenamefont {Khan}, \citenamefont {Iqbal},
  \citenamefont {Zhang},\ and\ \citenamefont {Guo}}]{khan2021novel}%
  \BibitemOpen
  \bibfield  {author} {\bibinfo {author} {\bibfnamefont {K.}~\bibnamefont
  {Khan}}, \bibinfo {author} {\bibfnamefont {A.~K.}\ \bibnamefont {Tareen}},
  \bibinfo {author} {\bibfnamefont {Q.~U.}\ \bibnamefont {Khan}}, \bibinfo
  {author} {\bibfnamefont {M.}~\bibnamefont {Iqbal}}, \bibinfo {author}
  {\bibfnamefont {H.}~\bibnamefont {Zhang}}, \ and\ \bibinfo {author}
  {\bibfnamefont {Z.}~\bibnamefont {Guo}},\ }\href {\doibase
  10.1039/D1QM00629K} {\bibfield  {journal} {\bibinfo  {journal} {Mater. Chem.
  Front.}\ }\textbf {\bibinfo {volume} {5}},\ \bibinfo {pages} {6333} (\bibinfo
  {year} {2021})}\BibitemShut {NoStop}%
\bibitem [{\citenamefont {Xia}\ \emph {et~al.}(2019)\citenamefont {Xia},
  \citenamefont {Wang}, \citenamefont {Hwang}, \citenamefont {Neto},\ and\
  \citenamefont {Yang}}]{xia2019black}%
  \BibitemOpen
  \bibfield  {author} {\bibinfo {author} {\bibfnamefont {F.}~\bibnamefont
  {Xia}}, \bibinfo {author} {\bibfnamefont {H.}~\bibnamefont {Wang}}, \bibinfo
  {author} {\bibfnamefont {J.~C.}\ \bibnamefont {Hwang}}, \bibinfo {author}
  {\bibfnamefont {A.~C.}\ \bibnamefont {Neto}}, \ and\ \bibinfo {author}
  {\bibfnamefont {L.}~\bibnamefont {Yang}},\ }\href
  {https://www.nature.com/articles/s42254-019-0043-5} {\bibfield  {journal}
  {\bibinfo  {journal} {Nat. Rev. Phys.}\ ,\ \bibinfo {pages} {1}} (\bibinfo
  {year} {2019})}\BibitemShut {NoStop}%
\bibitem [{\citenamefont {Li}\ \emph {et~al.}(2014)\citenamefont {Li},
  \citenamefont {Yu}, \citenamefont {Ye}, \citenamefont {Ge}, \citenamefont
  {Ou}, \citenamefont {Wu}, \citenamefont {Feng}, \citenamefont {Chen},\ and\
  \citenamefont {Zhang}}]{li2014black}%
  \BibitemOpen
  \bibfield  {author} {\bibinfo {author} {\bibfnamefont {L.}~\bibnamefont
  {Li}}, \bibinfo {author} {\bibfnamefont {Y.}~\bibnamefont {Yu}}, \bibinfo
  {author} {\bibfnamefont {G.~J.}\ \bibnamefont {Ye}}, \bibinfo {author}
  {\bibfnamefont {Q.}~\bibnamefont {Ge}}, \bibinfo {author} {\bibfnamefont
  {X.}~\bibnamefont {Ou}}, \bibinfo {author} {\bibfnamefont {H.}~\bibnamefont
  {Wu}}, \bibinfo {author} {\bibfnamefont {D.}~\bibnamefont {Feng}}, \bibinfo
  {author} {\bibfnamefont {X.~H.}\ \bibnamefont {Chen}}, \ and\ \bibinfo
  {author} {\bibfnamefont {Y.}~\bibnamefont {Zhang}},\ }\href
  {https://www.nature.com/articles/nnano.2014.35} {\bibfield  {journal}
  {\bibinfo  {journal} {Nature Nanotech.}\ }\textbf {\bibinfo {volume} {9}},\
  \bibinfo {pages} {372} (\bibinfo {year} {2014})}\BibitemShut {NoStop}%
\bibitem [{\citenamefont {Liu}\ \emph {et~al.}(2014)\citenamefont {Liu},
  \citenamefont {Neal}, \citenamefont {Zhu}, \citenamefont {Luo}, \citenamefont
  {Xu}, \citenamefont {Tom{\'a}nek},\ and\ \citenamefont
  {Ye}}]{liu2014phosphorene}%
  \BibitemOpen
  \bibfield  {author} {\bibinfo {author} {\bibfnamefont {H.}~\bibnamefont
  {Liu}}, \bibinfo {author} {\bibfnamefont {A.~T.}\ \bibnamefont {Neal}},
  \bibinfo {author} {\bibfnamefont {Z.}~\bibnamefont {Zhu}}, \bibinfo {author}
  {\bibfnamefont {Z.}~\bibnamefont {Luo}}, \bibinfo {author} {\bibfnamefont
  {X.}~\bibnamefont {Xu}}, \bibinfo {author} {\bibfnamefont {D.}~\bibnamefont
  {Tom{\'a}nek}}, \ and\ \bibinfo {author} {\bibfnamefont {P.~D.}\ \bibnamefont
  {Ye}},\ }\href {https://pubs.acs.org/doi/abs/10.1021/nn501226z} {\bibfield
  {journal} {\bibinfo  {journal} {ACS nano}\ }\textbf {\bibinfo {volume} {8}},\
  \bibinfo {pages} {4033} (\bibinfo {year} {2014})}\BibitemShut {NoStop}%
\bibitem [{\citenamefont {Zhu}\ and\ \citenamefont
  {Tom\'anek}(2014)}]{bluePPhysRevLett.112.176802}%
  \BibitemOpen
  \bibfield  {author} {\bibinfo {author} {\bibfnamefont {Z.}~\bibnamefont
  {Zhu}}\ and\ \bibinfo {author} {\bibfnamefont {D.}~\bibnamefont
  {Tom\'anek}},\ }\href {\doibase 10.1103/PhysRevLett.112.176802} {\bibfield
  {journal} {\bibinfo  {journal} {Phys. Rev. Lett.}\ }\textbf {\bibinfo
  {volume} {112}},\ \bibinfo {pages} {176802} (\bibinfo {year}
  {2014})}\BibitemShut {NoStop}%
\bibitem [{\citenamefont {Mardanya}\ \emph {et~al.}(2016)\citenamefont
  {Mardanya}, \citenamefont {Thakur}, \citenamefont {Bhowmick},\ and\
  \citenamefont {Agarwal}}]{mardanya2016four}%
  \BibitemOpen
  \bibfield  {author} {\bibinfo {author} {\bibfnamefont {S.}~\bibnamefont
  {Mardanya}}, \bibinfo {author} {\bibfnamefont {V.~K.}\ \bibnamefont
  {Thakur}}, \bibinfo {author} {\bibfnamefont {S.}~\bibnamefont {Bhowmick}}, \
  and\ \bibinfo {author} {\bibfnamefont {A.}~\bibnamefont {Agarwal}},\ }\href
  {https://journals.aps.org/prb/abstract/10.1103/PhysRevB.94.035423} {\bibfield
   {journal} {\bibinfo  {journal} {Phys. Rev. B}\ }\textbf {\bibinfo {volume}
  {94}},\ \bibinfo {pages} {035423} (\bibinfo {year} {2016})}\BibitemShut
  {NoStop}%
\bibitem [{\citenamefont {Zhong}\ \emph {et~al.}(2018)\citenamefont {Zhong},
  \citenamefont {Xia}, \citenamefont {Pan}, \citenamefont {Liu}, \citenamefont
  {Chen}, \citenamefont {Deng}, \citenamefont {Li},\ and\ \citenamefont
  {Wei}}]{zhong2018thickness}%
  \BibitemOpen
  \bibfield  {author} {\bibinfo {author} {\bibfnamefont {M.}~\bibnamefont
  {Zhong}}, \bibinfo {author} {\bibfnamefont {Q.}~\bibnamefont {Xia}}, \bibinfo
  {author} {\bibfnamefont {L.}~\bibnamefont {Pan}}, \bibinfo {author}
  {\bibfnamefont {Y.}~\bibnamefont {Liu}}, \bibinfo {author} {\bibfnamefont
  {Y.}~\bibnamefont {Chen}}, \bibinfo {author} {\bibfnamefont {H.-X.}\
  \bibnamefont {Deng}}, \bibinfo {author} {\bibfnamefont {J.}~\bibnamefont
  {Li}}, \ and\ \bibinfo {author} {\bibfnamefont {Z.}~\bibnamefont {Wei}},\
  }\href {https://onlinelibrary.wiley.com/doi/full/10.1002/adfm.201802581}
  {\bibfield  {journal} {\bibinfo  {journal} {Adv. Funct. Mater.}\ }\textbf
  {\bibinfo {volume} {28}},\ \bibinfo {pages} {1802581} (\bibinfo {year}
  {2018})}\BibitemShut {NoStop}%
\bibitem [{\citenamefont {Kamal}\ and\ \citenamefont
  {Ezawa}(2015)}]{kamal2015arsenene}%
  \BibitemOpen
  \bibfield  {author} {\bibinfo {author} {\bibfnamefont {C.}~\bibnamefont
  {Kamal}}\ and\ \bibinfo {author} {\bibfnamefont {M.}~\bibnamefont {Ezawa}},\
  }\href {https://journals.aps.org/prb/abstract/10.1103/PhysRevB.91.085423}
  {\bibfield  {journal} {\bibinfo  {journal} {Phys. Rev. B}\ }\textbf {\bibinfo
  {volume} {91}},\ \bibinfo {pages} {085423} (\bibinfo {year}
  {2015})}\BibitemShut {NoStop}%
\bibitem [{\citenamefont {Shao}\ \emph {et~al.}(2018)\citenamefont {Shao},
  \citenamefont {Liu}, \citenamefont {Cheng}, \citenamefont {Wu}, \citenamefont
  {Liu}, \citenamefont {Liu}, \citenamefont {Wang}, \citenamefont {Zhu},
  \citenamefont {Wang}, \citenamefont {Shi} \emph
  {et~al.}}]{shao2018epitaxial}%
  \BibitemOpen
  \bibfield  {author} {\bibinfo {author} {\bibfnamefont {Y.}~\bibnamefont
  {Shao}}, \bibinfo {author} {\bibfnamefont {Z.-L.}\ \bibnamefont {Liu}},
  \bibinfo {author} {\bibfnamefont {C.}~\bibnamefont {Cheng}}, \bibinfo
  {author} {\bibfnamefont {X.}~\bibnamefont {Wu}}, \bibinfo {author}
  {\bibfnamefont {H.}~\bibnamefont {Liu}}, \bibinfo {author} {\bibfnamefont
  {C.}~\bibnamefont {Liu}}, \bibinfo {author} {\bibfnamefont {J.-O.}\
  \bibnamefont {Wang}}, \bibinfo {author} {\bibfnamefont {S.-Y.}\ \bibnamefont
  {Zhu}}, \bibinfo {author} {\bibfnamefont {Y.-Q.}\ \bibnamefont {Wang}},
  \bibinfo {author} {\bibfnamefont {D.-X.}\ \bibnamefont {Shi}},  \emph
  {et~al.},\ }\href {https://pubs.acs.org/doi/abs/10.1021/acs.nanolett.8b00429}
  {\bibfield  {journal} {\bibinfo  {journal} {Nano Lett.}\ }\textbf {\bibinfo
  {volume} {18}},\ \bibinfo {pages} {2133} (\bibinfo {year}
  {2018})}\BibitemShut {NoStop}%
\bibitem [{\citenamefont {Wu}\ \emph {et~al.}(2017)\citenamefont {Wu},
  \citenamefont {Shao}, \citenamefont {Liu}, \citenamefont {Feng},
  \citenamefont {Wang}, \citenamefont {Sun}, \citenamefont {Liu}, \citenamefont
  {Wang}, \citenamefont {Liu}, \citenamefont {Zhu} \emph
  {et~al.}}]{wu2017epitaxial}%
  \BibitemOpen
  \bibfield  {author} {\bibinfo {author} {\bibfnamefont {X.}~\bibnamefont
  {Wu}}, \bibinfo {author} {\bibfnamefont {Y.}~\bibnamefont {Shao}}, \bibinfo
  {author} {\bibfnamefont {H.}~\bibnamefont {Liu}}, \bibinfo {author}
  {\bibfnamefont {Z.}~\bibnamefont {Feng}}, \bibinfo {author} {\bibfnamefont
  {Y.-L.}\ \bibnamefont {Wang}}, \bibinfo {author} {\bibfnamefont {J.-T.}\
  \bibnamefont {Sun}}, \bibinfo {author} {\bibfnamefont {C.}~\bibnamefont
  {Liu}}, \bibinfo {author} {\bibfnamefont {J.-O.}\ \bibnamefont {Wang}},
  \bibinfo {author} {\bibfnamefont {Z.-L.}\ \bibnamefont {Liu}}, \bibinfo
  {author} {\bibfnamefont {S.-Y.}\ \bibnamefont {Zhu}},  \emph {et~al.},\
  }\href {https://onlinelibrary.wiley.com/doi/full/10.1002/adma.201605407}
  {\bibfield  {journal} {\bibinfo  {journal} {Adv. Mater.}\ }\textbf {\bibinfo
  {volume} {29}},\ \bibinfo {pages} {1605407} (\bibinfo {year}
  {2017})}\BibitemShut {NoStop}%
\bibitem [{\citenamefont {Wang}\ \emph {et~al.}(2015)\citenamefont {Wang},
  \citenamefont {Pandey},\ and\ \citenamefont {Karna}}]{wang2015atomically}%
  \BibitemOpen
  \bibfield  {author} {\bibinfo {author} {\bibfnamefont {G.}~\bibnamefont
  {Wang}}, \bibinfo {author} {\bibfnamefont {R.}~\bibnamefont {Pandey}}, \ and\
  \bibinfo {author} {\bibfnamefont {S.~P.}\ \bibnamefont {Karna}},\ }\href
  {https://pubs.acs.org/doi/abs/10.1021/acsami.5b02441} {\bibfield  {journal}
  {\bibinfo  {journal} {ACS Appl. Mater. Inter.}\ }\textbf {\bibinfo {volume}
  {7}},\ \bibinfo {pages} {11490} (\bibinfo {year} {2015})}\BibitemShut
  {NoStop}%
\bibitem [{\citenamefont {Lu}\ \emph {et~al.}(2014)\citenamefont {Lu},
  \citenamefont {Xu}, \citenamefont {Zeng}, \citenamefont {Yao}, \citenamefont
  {Shen}, \citenamefont {Yang}, \citenamefont {Luo}, \citenamefont {Pan},
  \citenamefont {Wu}, \citenamefont {Das} \emph {et~al.}}]{lu2014topological}%
  \BibitemOpen
  \bibfield  {author} {\bibinfo {author} {\bibfnamefont {Y.}~\bibnamefont
  {Lu}}, \bibinfo {author} {\bibfnamefont {W.}~\bibnamefont {Xu}}, \bibinfo
  {author} {\bibfnamefont {M.}~\bibnamefont {Zeng}}, \bibinfo {author}
  {\bibfnamefont {G.}~\bibnamefont {Yao}}, \bibinfo {author} {\bibfnamefont
  {L.}~\bibnamefont {Shen}}, \bibinfo {author} {\bibfnamefont {M.}~\bibnamefont
  {Yang}}, \bibinfo {author} {\bibfnamefont {Z.}~\bibnamefont {Luo}}, \bibinfo
  {author} {\bibfnamefont {F.}~\bibnamefont {Pan}}, \bibinfo {author}
  {\bibfnamefont {K.}~\bibnamefont {Wu}}, \bibinfo {author} {\bibfnamefont
  {T.}~\bibnamefont {Das}},  \emph {et~al.},\ }\href
  {https://pubs.acs.org/doi/abs/10.1021/nl502997v} {\bibfield  {journal}
  {\bibinfo  {journal} {Nano Lett.}\ }\textbf {\bibinfo {volume} {15}},\
  \bibinfo {pages} {80} (\bibinfo {year} {2014})}\BibitemShut {NoStop}%
\bibitem [{\citenamefont {Liu}\ \emph {et~al.}(2011)\citenamefont {Liu},
  \citenamefont {Liu}, \citenamefont {Wu}, \citenamefont {Duan}, \citenamefont
  {Liu},\ and\ \citenamefont {Wu}}]{liu2011stable}%
  \BibitemOpen
  \bibfield  {author} {\bibinfo {author} {\bibfnamefont {Z.}~\bibnamefont
  {Liu}}, \bibinfo {author} {\bibfnamefont {C.-X.}\ \bibnamefont {Liu}},
  \bibinfo {author} {\bibfnamefont {Y.-S.}\ \bibnamefont {Wu}}, \bibinfo
  {author} {\bibfnamefont {W.-H.}\ \bibnamefont {Duan}}, \bibinfo {author}
  {\bibfnamefont {F.}~\bibnamefont {Liu}}, \ and\ \bibinfo {author}
  {\bibfnamefont {J.}~\bibnamefont {Wu}},\ }\href
  {https://journals.aps.org/prl/abstract/10.1103/PhysRevLett.107.136805}
  {\bibfield  {journal} {\bibinfo  {journal} {Phys. Rev. Lett.}\ }\textbf
  {\bibinfo {volume} {107}},\ \bibinfo {pages} {136805} (\bibinfo {year}
  {2011})}\BibitemShut {NoStop}%
\bibitem [{\citenamefont {Drozdov}\ \emph {et~al.}(2014)\citenamefont
  {Drozdov}, \citenamefont {Alexandradinata}, \citenamefont {Jeon},
  \citenamefont {Nadj-Perge}, \citenamefont {Ji}, \citenamefont {Cava},
  \citenamefont {Bernevig},\ and\ \citenamefont {Yazdani}}]{drozdov2014one}%
  \BibitemOpen
  \bibfield  {author} {\bibinfo {author} {\bibfnamefont {I.~K.}\ \bibnamefont
  {Drozdov}}, \bibinfo {author} {\bibfnamefont {A.}~\bibnamefont
  {Alexandradinata}}, \bibinfo {author} {\bibfnamefont {S.}~\bibnamefont
  {Jeon}}, \bibinfo {author} {\bibfnamefont {S.}~\bibnamefont {Nadj-Perge}},
  \bibinfo {author} {\bibfnamefont {H.}~\bibnamefont {Ji}}, \bibinfo {author}
  {\bibfnamefont {R.~J.}\ \bibnamefont {Cava}}, \bibinfo {author}
  {\bibfnamefont {B.~A.}\ \bibnamefont {Bernevig}}, \ and\ \bibinfo {author}
  {\bibfnamefont {A.}~\bibnamefont {Yazdani}},\ }\href
  {https://www.nature.com/articles/nphys3048} {\bibfield  {journal} {\bibinfo
  {journal} {Nat. Phys.}\ }\textbf {\bibinfo {volume} {10}},\ \bibinfo {pages}
  {664} (\bibinfo {year} {2014})}\BibitemShut {NoStop}%
\bibitem [{\citenamefont {Zhang}\ \emph {et~al.}(2016)\citenamefont {Zhang},
  \citenamefont {Xie}, \citenamefont {Li}, \citenamefont {Yan}, \citenamefont
  {Li}, \citenamefont {Kan}, \citenamefont {Liu}, \citenamefont {Chen},\ and\
  \citenamefont {Zeng}}]{zhang2016semiconducting}%
  \BibitemOpen
  \bibfield  {author} {\bibinfo {author} {\bibfnamefont {S.}~\bibnamefont
  {Zhang}}, \bibinfo {author} {\bibfnamefont {M.}~\bibnamefont {Xie}}, \bibinfo
  {author} {\bibfnamefont {F.}~\bibnamefont {Li}}, \bibinfo {author}
  {\bibfnamefont {Z.}~\bibnamefont {Yan}}, \bibinfo {author} {\bibfnamefont
  {Y.}~\bibnamefont {Li}}, \bibinfo {author} {\bibfnamefont {E.}~\bibnamefont
  {Kan}}, \bibinfo {author} {\bibfnamefont {W.}~\bibnamefont {Liu}}, \bibinfo
  {author} {\bibfnamefont {Z.}~\bibnamefont {Chen}}, \ and\ \bibinfo {author}
  {\bibfnamefont {H.}~\bibnamefont {Zeng}},\ }\href
  {https://onlinelibrary.wiley.com/doi/full/10.1002/anie.201507568} {\bibfield
  {journal} {\bibinfo  {journal} {Angew. Chem.-Int. Edit}\ }\textbf {\bibinfo
  {volume} {55}},\ \bibinfo {pages} {1666} (\bibinfo {year}
  {2016})}\BibitemShut {NoStop}%
\bibitem [{\citenamefont {Qiao}\ \emph {et~al.}(2014)\citenamefont {Qiao},
  \citenamefont {Kong}, \citenamefont {Hu}, \citenamefont {Yang},\ and\
  \citenamefont {Ji}}]{qiao2014high}%
  \BibitemOpen
  \bibfield  {author} {\bibinfo {author} {\bibfnamefont {J.}~\bibnamefont
  {Qiao}}, \bibinfo {author} {\bibfnamefont {X.}~\bibnamefont {Kong}}, \bibinfo
  {author} {\bibfnamefont {Z.-X.}\ \bibnamefont {Hu}}, \bibinfo {author}
  {\bibfnamefont {F.}~\bibnamefont {Yang}}, \ and\ \bibinfo {author}
  {\bibfnamefont {W.}~\bibnamefont {Ji}},\ }\href {\doibase 10.1038/ncomms5475}
  {\bibfield  {journal} {\bibinfo  {journal} {Nat. Commun.}\ }\textbf {\bibinfo
  {volume} {5}},\ \bibinfo {pages} {4475} (\bibinfo {year} {2014})}\BibitemShut
  {NoStop}%
\bibitem [{\citenamefont {Xia}\ \emph {et~al.}(2014)\citenamefont {Xia},
  \citenamefont {Wang},\ and\ \citenamefont {Jia}}]{xia2014rediscovering}%
  \BibitemOpen
  \bibfield  {author} {\bibinfo {author} {\bibfnamefont {F.}~\bibnamefont
  {Xia}}, \bibinfo {author} {\bibfnamefont {H.}~\bibnamefont {Wang}}, \ and\
  \bibinfo {author} {\bibfnamefont {Y.}~\bibnamefont {Jia}},\ }\href
  {https://www.nature.com/articles/ncomms5458} {\bibfield  {journal} {\bibinfo
  {journal} {Nat. Commun.}\ }\textbf {\bibinfo {volume} {5}},\ \bibinfo {pages}
  {4458} (\bibinfo {year} {2014})}\BibitemShut {NoStop}%
\bibitem [{\citenamefont {Frost}\ \emph {et~al.}(2016)\citenamefont {Frost},
  \citenamefont {Howie}, \citenamefont {Dalladay-Simpson}, \citenamefont
  {Goncharov},\ and\ \citenamefont {Gregoryanz}}]{researchnitrogen}%
  \BibitemOpen
  \bibfield  {author} {\bibinfo {author} {\bibfnamefont {M.}~\bibnamefont
  {Frost}}, \bibinfo {author} {\bibfnamefont {R.~T.}\ \bibnamefont {Howie}},
  \bibinfo {author} {\bibfnamefont {P.}~\bibnamefont {Dalladay-Simpson}},
  \bibinfo {author} {\bibfnamefont {A.~F.}\ \bibnamefont {Goncharov}}, \ and\
  \bibinfo {author} {\bibfnamefont {E.}~\bibnamefont {Gregoryanz}},\ }\href
  {\doibase 10.1103/PhysRevB.93.024113} {\bibfield  {journal} {\bibinfo
  {journal} {Phys. Rev. B}\ }\textbf {\bibinfo {volume} {93}},\ \bibinfo
  {pages} {024113} (\bibinfo {year} {2016})}\BibitemShut {NoStop}%
\bibitem [{\citenamefont {Lin}\ \emph {et~al.}(2018)\citenamefont {Lin},
  \citenamefont {Choi}, \citenamefont {Zhang}, \citenamefont {Qin},
  \citenamefont {Yi}, \citenamefont {Wang}, \citenamefont {Li}, \citenamefont
  {Wang}, \citenamefont {Zhang}, \citenamefont {Sun}, \citenamefont {Wei},
  \citenamefont {Zhang}, \citenamefont {Guo}, \citenamefont {Lu}, \citenamefont
  {Cho}, \citenamefont {Zeng},\ and\ \citenamefont
  {Zhang}}]{PhysRevLett.121.096401}%
  \BibitemOpen
  \bibfield  {author} {\bibinfo {author} {\bibfnamefont {Z.}~\bibnamefont
  {Lin}}, \bibinfo {author} {\bibfnamefont {J.-H.}\ \bibnamefont {Choi}},
  \bibinfo {author} {\bibfnamefont {Q.}~\bibnamefont {Zhang}}, \bibinfo
  {author} {\bibfnamefont {W.}~\bibnamefont {Qin}}, \bibinfo {author}
  {\bibfnamefont {S.}~\bibnamefont {Yi}}, \bibinfo {author} {\bibfnamefont
  {P.}~\bibnamefont {Wang}}, \bibinfo {author} {\bibfnamefont {L.}~\bibnamefont
  {Li}}, \bibinfo {author} {\bibfnamefont {Y.}~\bibnamefont {Wang}}, \bibinfo
  {author} {\bibfnamefont {H.}~\bibnamefont {Zhang}}, \bibinfo {author}
  {\bibfnamefont {Z.}~\bibnamefont {Sun}}, \bibinfo {author} {\bibfnamefont
  {L.}~\bibnamefont {Wei}}, \bibinfo {author} {\bibfnamefont {S.}~\bibnamefont
  {Zhang}}, \bibinfo {author} {\bibfnamefont {T.}~\bibnamefont {Guo}}, \bibinfo
  {author} {\bibfnamefont {Q.}~\bibnamefont {Lu}}, \bibinfo {author}
  {\bibfnamefont {J.-H.}\ \bibnamefont {Cho}}, \bibinfo {author} {\bibfnamefont
  {C.}~\bibnamefont {Zeng}}, \ and\ \bibinfo {author} {\bibfnamefont
  {Z.}~\bibnamefont {Zhang}},\ }\href {\doibase 10.1103/PhysRevLett.121.096401}
  {\bibfield  {journal} {\bibinfo  {journal} {Phys. Rev. Lett.}\ }\textbf
  {\bibinfo {volume} {121}},\ \bibinfo {pages} {096401} (\bibinfo {year}
  {2018})}\BibitemShut {NoStop}%
\bibitem [{\citenamefont {Liu}\ \emph {et~al.}(2020)\citenamefont {Liu},
  \citenamefont {Li}, \citenamefont {Wang}, \citenamefont {Wang}, \citenamefont
  {Wen}, \citenamefont {Jiang}, \citenamefont {Lu}, \citenamefont {Yan},
  \citenamefont {Huang}, \citenamefont {Shen} \emph {et~al.}}]{flatdirac}%
  \BibitemOpen
  \bibfield  {author} {\bibinfo {author} {\bibfnamefont {Z.}~\bibnamefont
  {Liu}}, \bibinfo {author} {\bibfnamefont {M.}~\bibnamefont {Li}}, \bibinfo
  {author} {\bibfnamefont {Q.}~\bibnamefont {Wang}}, \bibinfo {author}
  {\bibfnamefont {G.}~\bibnamefont {Wang}}, \bibinfo {author} {\bibfnamefont
  {C.}~\bibnamefont {Wen}}, \bibinfo {author} {\bibfnamefont {K.}~\bibnamefont
  {Jiang}}, \bibinfo {author} {\bibfnamefont {X.}~\bibnamefont {Lu}}, \bibinfo
  {author} {\bibfnamefont {S.}~\bibnamefont {Yan}}, \bibinfo {author}
  {\bibfnamefont {Y.}~\bibnamefont {Huang}}, \bibinfo {author} {\bibfnamefont
  {D.}~\bibnamefont {Shen}},  \emph {et~al.},\ }\href {\doibase
  10.1038/s41467-020-17462-4} {\bibfield  {journal} {\bibinfo  {journal} {Nat.
  Commun.}\ }\textbf {\bibinfo {volume} {11}},\ \bibinfo {pages} {4002}
  (\bibinfo {year} {2020})}\BibitemShut {NoStop}%
\bibitem [{\citenamefont {Kang}\ \emph {et~al.}(2020)\citenamefont {Kang},
  \citenamefont {Ye}, \citenamefont {Fang}, \citenamefont {You}, \citenamefont
  {Levitan}, \citenamefont {Han}, \citenamefont {Facio}, \citenamefont
  {Jozwiak}, \citenamefont {Bostwick}, \citenamefont {Rotenberg} \emph
  {et~al.}}]{kang2020dirac}%
  \BibitemOpen
  \bibfield  {author} {\bibinfo {author} {\bibfnamefont {M.}~\bibnamefont
  {Kang}}, \bibinfo {author} {\bibfnamefont {L.}~\bibnamefont {Ye}}, \bibinfo
  {author} {\bibfnamefont {S.}~\bibnamefont {Fang}}, \bibinfo {author}
  {\bibfnamefont {J.-S.}\ \bibnamefont {You}}, \bibinfo {author} {\bibfnamefont
  {A.}~\bibnamefont {Levitan}}, \bibinfo {author} {\bibfnamefont
  {M.}~\bibnamefont {Han}}, \bibinfo {author} {\bibfnamefont {J.~I.}\
  \bibnamefont {Facio}}, \bibinfo {author} {\bibfnamefont {C.}~\bibnamefont
  {Jozwiak}}, \bibinfo {author} {\bibfnamefont {A.}~\bibnamefont {Bostwick}},
  \bibinfo {author} {\bibfnamefont {E.}~\bibnamefont {Rotenberg}},  \emph
  {et~al.},\ }\href {\doibase 10.1038/s41563-019-0531-0} {\bibfield  {journal}
  {\bibinfo  {journal} {Nat. Mater.}\ }\textbf {\bibinfo {volume} {19}},\
  \bibinfo {pages} {163} (\bibinfo {year} {2020})}\BibitemShut {NoStop}%
\bibitem [{\citenamefont {Yin}\ \emph {et~al.}(2018)\citenamefont {Yin},
  \citenamefont {Zhang}, \citenamefont {Li}, \citenamefont {Jiang},
  \citenamefont {Chang}, \citenamefont {Zhang}, \citenamefont {Lian},
  \citenamefont {Xiang}, \citenamefont {Belopolski}, \citenamefont {Zheng}
  \emph {et~al.}}]{yin2018giant}%
  \BibitemOpen
  \bibfield  {author} {\bibinfo {author} {\bibfnamefont {J.-X.}\ \bibnamefont
  {Yin}}, \bibinfo {author} {\bibfnamefont {S.~S.}\ \bibnamefont {Zhang}},
  \bibinfo {author} {\bibfnamefont {H.}~\bibnamefont {Li}}, \bibinfo {author}
  {\bibfnamefont {K.}~\bibnamefont {Jiang}}, \bibinfo {author} {\bibfnamefont
  {G.}~\bibnamefont {Chang}}, \bibinfo {author} {\bibfnamefont
  {B.}~\bibnamefont {Zhang}}, \bibinfo {author} {\bibfnamefont
  {B.}~\bibnamefont {Lian}}, \bibinfo {author} {\bibfnamefont {C.}~\bibnamefont
  {Xiang}}, \bibinfo {author} {\bibfnamefont {I.}~\bibnamefont {Belopolski}},
  \bibinfo {author} {\bibfnamefont {H.}~\bibnamefont {Zheng}},  \emph
  {et~al.},\ }\href {\doibase 10.1038/s41586-018-0502-7} {\bibfield  {journal}
  {\bibinfo  {journal} {Nature}\ }\textbf {\bibinfo {volume} {562}},\ \bibinfo
  {pages} {91} (\bibinfo {year} {2018})}\BibitemShut {NoStop}%
\bibitem [{\citenamefont {Zhang}\ \emph {et~al.}(2020)\citenamefont {Zhang},
  \citenamefont {Yin}, \citenamefont {Ikhlas}, \citenamefont {Tien},
  \citenamefont {Wang}, \citenamefont {Shumiya}, \citenamefont {Chang},
  \citenamefont {Tsirkin}, \citenamefont {Shi}, \citenamefont {Yi},
  \citenamefont {Guguchia}, \citenamefont {Li}, \citenamefont {Wang},
  \citenamefont {Chang}, \citenamefont {Wang}, \citenamefont {Yang},
  \citenamefont {Neupert}, \citenamefont {Nakatsuji},\ and\ \citenamefont
  {Hasan}}]{manybodyeffect}%
  \BibitemOpen
  \bibfield  {author} {\bibinfo {author} {\bibfnamefont {S.~S.}\ \bibnamefont
  {Zhang}}, \bibinfo {author} {\bibfnamefont {J.-X.}\ \bibnamefont {Yin}},
  \bibinfo {author} {\bibfnamefont {M.}~\bibnamefont {Ikhlas}}, \bibinfo
  {author} {\bibfnamefont {H.-J.}\ \bibnamefont {Tien}}, \bibinfo {author}
  {\bibfnamefont {R.}~\bibnamefont {Wang}}, \bibinfo {author} {\bibfnamefont
  {N.}~\bibnamefont {Shumiya}}, \bibinfo {author} {\bibfnamefont
  {G.}~\bibnamefont {Chang}}, \bibinfo {author} {\bibfnamefont {S.~S.}\
  \bibnamefont {Tsirkin}}, \bibinfo {author} {\bibfnamefont {Y.}~\bibnamefont
  {Shi}}, \bibinfo {author} {\bibfnamefont {C.}~\bibnamefont {Yi}}, \bibinfo
  {author} {\bibfnamefont {Z.}~\bibnamefont {Guguchia}}, \bibinfo {author}
  {\bibfnamefont {H.}~\bibnamefont {Li}}, \bibinfo {author} {\bibfnamefont
  {W.}~\bibnamefont {Wang}}, \bibinfo {author} {\bibfnamefont {T.-R.}\
  \bibnamefont {Chang}}, \bibinfo {author} {\bibfnamefont {Z.}~\bibnamefont
  {Wang}}, \bibinfo {author} {\bibfnamefont {Y.-f.}\ \bibnamefont {Yang}},
  \bibinfo {author} {\bibfnamefont {T.}~\bibnamefont {Neupert}}, \bibinfo
  {author} {\bibfnamefont {S.}~\bibnamefont {Nakatsuji}}, \ and\ \bibinfo
  {author} {\bibfnamefont {M.~Z.}\ \bibnamefont {Hasan}},\ }\href {\doibase
  10.1103/PhysRevLett.125.046401} {\bibfield  {journal} {\bibinfo  {journal}
  {Phys. Rev. Lett.}\ }\textbf {\bibinfo {volume} {125}},\ \bibinfo {pages}
  {046401} (\bibinfo {year} {2020})}\BibitemShut {NoStop}%
\bibitem [{\citenamefont {Yin}\ \emph {et~al.}(2022)\citenamefont {Yin},
  \citenamefont {Lian},\ and\ \citenamefont {Hasan}}]{yin2022topological}%
  \BibitemOpen
  \bibfield  {author} {\bibinfo {author} {\bibfnamefont {J.-X.}\ \bibnamefont
  {Yin}}, \bibinfo {author} {\bibfnamefont {B.}~\bibnamefont {Lian}}, \ and\
  \bibinfo {author} {\bibfnamefont {M.~Z.}\ \bibnamefont {Hasan}},\ }\href
  {\doibase 10.1038/s41586-022-05516-0} {\bibfield  {journal} {\bibinfo
  {journal} {Nature}\ }\textbf {\bibinfo {volume} {612}},\ \bibinfo {pages}
  {647} (\bibinfo {year} {2022})}\BibitemShut {NoStop}%
\bibitem [{\citenamefont {Kresse}\ and\ \citenamefont
  {Hafner}(1993)}]{PhysRevB.47.558}%
  \BibitemOpen
  \bibfield  {author} {\bibinfo {author} {\bibfnamefont {G.}~\bibnamefont
  {Kresse}}\ and\ \bibinfo {author} {\bibfnamefont {J.}~\bibnamefont
  {Hafner}},\ }\href {\doibase 10.1103/PhysRevB.47.558} {\bibfield  {journal}
  {\bibinfo  {journal} {Phys. Rev. B}\ }\textbf {\bibinfo {volume} {47}},\
  \bibinfo {pages} {558} (\bibinfo {year} {1993})}\BibitemShut {NoStop}%
\bibitem [{\citenamefont {Kresse}\ and\ \citenamefont
  {Hafner}(1994)}]{PhysRevB.49.14251}%
  \BibitemOpen
  \bibfield  {author} {\bibinfo {author} {\bibfnamefont {G.}~\bibnamefont
  {Kresse}}\ and\ \bibinfo {author} {\bibfnamefont {J.}~\bibnamefont
  {Hafner}},\ }\href {\doibase 10.1103/PhysRevB.49.14251} {\bibfield  {journal}
  {\bibinfo  {journal} {Phys. Rev. B}\ }\textbf {\bibinfo {volume} {49}},\
  \bibinfo {pages} {14251} (\bibinfo {year} {1994})}\BibitemShut {NoStop}%
\bibitem [{\citenamefont {Kresse}\ and\ \citenamefont
  {Furthm{\"u}ller}(1996)}]{kresse1996efficiency}%
  \BibitemOpen
  \bibfield  {author} {\bibinfo {author} {\bibfnamefont {G.}~\bibnamefont
  {Kresse}}\ and\ \bibinfo {author} {\bibfnamefont {J.}~\bibnamefont
  {Furthm{\"u}ller}},\ }\href {\doibase 10.1016/0927-0256(96)00008-0}
  {\bibfield  {journal} {\bibinfo  {journal} {Comp. Mater. Sci.}\ }\textbf
  {\bibinfo {volume} {6}},\ \bibinfo {pages} {15} (\bibinfo {year}
  {1996})}\BibitemShut {NoStop}%
\bibitem [{\citenamefont {Kresse}\ and\ \citenamefont
  {Furthm\"uller}(1996)}]{PhysRevB.54.11169}%
  \BibitemOpen
  \bibfield  {author} {\bibinfo {author} {\bibfnamefont {G.}~\bibnamefont
  {Kresse}}\ and\ \bibinfo {author} {\bibfnamefont {J.}~\bibnamefont
  {Furthm\"uller}},\ }\href {\doibase 10.1103/PhysRevB.54.11169} {\bibfield
  {journal} {\bibinfo  {journal} {Phys. Rev. B}\ }\textbf {\bibinfo {volume}
  {54}},\ \bibinfo {pages} {11169} (\bibinfo {year} {1996})}\BibitemShut
  {NoStop}%
\bibitem [{\citenamefont {Bl\"ochl}(1994)}]{PhysRevB.50.17953}%
  \BibitemOpen
  \bibfield  {author} {\bibinfo {author} {\bibfnamefont {P.~E.}\ \bibnamefont
  {Bl\"ochl}},\ }\href {\doibase 10.1103/PhysRevB.50.17953} {\bibfield
  {journal} {\bibinfo  {journal} {Phys. Rev. B}\ }\textbf {\bibinfo {volume}
  {50}},\ \bibinfo {pages} {17953} (\bibinfo {year} {1994})}\BibitemShut
  {NoStop}%
\bibitem [{\citenamefont {Perdew}\ \emph {et~al.}(1996)\citenamefont {Perdew},
  \citenamefont {Burke},\ and\ \citenamefont
  {Ernzerhof}}]{PhysRevLett.77.3865}%
  \BibitemOpen
  \bibfield  {author} {\bibinfo {author} {\bibfnamefont {J.~P.}\ \bibnamefont
  {Perdew}}, \bibinfo {author} {\bibfnamefont {K.}~\bibnamefont {Burke}}, \
  and\ \bibinfo {author} {\bibfnamefont {M.}~\bibnamefont {Ernzerhof}},\ }\href
  {\doibase 10.1103/PhysRevLett.77.3865} {\bibfield  {journal} {\bibinfo
  {journal} {Phys. Rev. Lett.}\ }\textbf {\bibinfo {volume} {77}},\ \bibinfo
  {pages} {3865} (\bibinfo {year} {1996})}\BibitemShut {NoStop}%
\bibitem [{\citenamefont {Krukau}\ \emph {et~al.}(2006)\citenamefont {Krukau},
  \citenamefont {Vydrov}, \citenamefont {Izmaylov},\ and\ \citenamefont
  {Scuseria}}]{krukau2006influence}%
  \BibitemOpen
  \bibfield  {author} {\bibinfo {author} {\bibfnamefont {A.~V.}\ \bibnamefont
  {Krukau}}, \bibinfo {author} {\bibfnamefont {O.~A.}\ \bibnamefont {Vydrov}},
  \bibinfo {author} {\bibfnamefont {A.~F.}\ \bibnamefont {Izmaylov}}, \ and\
  \bibinfo {author} {\bibfnamefont {G.~E.}\ \bibnamefont {Scuseria}},\ }\href
  {\doibase 10.1063/1.2404663} {\bibfield  {journal} {\bibinfo  {journal} {J.
  Phys. Chem. C}\ }\textbf {\bibinfo {volume} {125}} (\bibinfo {year} {2006}),\
  10.1063/1.2404663}\BibitemShut {NoStop}%
\bibitem [{\citenamefont {Monkhorst}\ and\ \citenamefont
  {Pack}(1976)}]{PhysRevB.13.5188}%
  \BibitemOpen
  \bibfield  {author} {\bibinfo {author} {\bibfnamefont {H.~J.}\ \bibnamefont
  {Monkhorst}}\ and\ \bibinfo {author} {\bibfnamefont {J.~D.}\ \bibnamefont
  {Pack}},\ }\href {\doibase 10.1103/PhysRevB.13.5188} {\bibfield  {journal}
  {\bibinfo  {journal} {Phys. Rev. B}\ }\textbf {\bibinfo {volume} {13}},\
  \bibinfo {pages} {5188} (\bibinfo {year} {1976})}\BibitemShut {NoStop}%
\bibitem [{\citenamefont {Togo}\ and\ \citenamefont
  {Tanaka}(2015)}]{togo2015first}%
  \BibitemOpen
  \bibfield  {author} {\bibinfo {author} {\bibfnamefont {A.}~\bibnamefont
  {Togo}}\ and\ \bibinfo {author} {\bibfnamefont {I.}~\bibnamefont {Tanaka}},\
  }\href {\doibase 10.1016/j.scriptamat.2015.07.021} {\bibfield  {journal}
  {\bibinfo  {journal} {Scripta Mater.}\ }\textbf {\bibinfo {volume} {108}},\
  \bibinfo {pages} {1} (\bibinfo {year} {2015})}\BibitemShut {NoStop}%
\bibitem [{\citenamefont {Zhang}\ \emph {et~al.}(2023)\citenamefont {Zhang},
  \citenamefont {Yu}, \citenamefont {Liu}, \citenamefont {Li}, \citenamefont
  {Yang},\ and\ \citenamefont {Yao}}]{zhang2023magnetickp}%
  \BibitemOpen
  \bibfield  {author} {\bibinfo {author} {\bibfnamefont {Z.}~\bibnamefont
  {Zhang}}, \bibinfo {author} {\bibfnamefont {Z.-M.}\ \bibnamefont {Yu}},
  \bibinfo {author} {\bibfnamefont {G.-B.}\ \bibnamefont {Liu}}, \bibinfo
  {author} {\bibfnamefont {Z.}~\bibnamefont {Li}}, \bibinfo {author}
  {\bibfnamefont {S.~A.}\ \bibnamefont {Yang}}, \ and\ \bibinfo {author}
  {\bibfnamefont {Y.}~\bibnamefont {Yao}},\ }\href {\doibase
  https://doi.org/10.1016/j.cpc.2023.108784} {\bibfield  {journal} {\bibinfo
  {journal} {Comput. Phys. Commun.}\ }\textbf {\bibinfo {volume} {290}},\
  \bibinfo {pages} {108784} (\bibinfo {year} {2023})}\BibitemShut {NoStop}%
\bibitem [{\citenamefont {Wu}\ \emph {et~al.}(2018)\citenamefont {Wu},
  \citenamefont {Zhang}, \citenamefont {Song}, \citenamefont {Troyer},\ and\
  \citenamefont {Soluyanov}}]{wanniertool2017}%
  \BibitemOpen
  \bibfield  {author} {\bibinfo {author} {\bibfnamefont {Q.}~\bibnamefont
  {Wu}}, \bibinfo {author} {\bibfnamefont {S.}~\bibnamefont {Zhang}}, \bibinfo
  {author} {\bibfnamefont {H.-F.}\ \bibnamefont {Song}}, \bibinfo {author}
  {\bibfnamefont {M.}~\bibnamefont {Troyer}}, \ and\ \bibinfo {author}
  {\bibfnamefont {A.~A.}\ \bibnamefont {Soluyanov}},\ }\href {\doibase
  https://doi.org/10.1016/j.cpc.2017.09.033} {\bibfield  {journal} {\bibinfo
  {journal} {Comp. Phys. Commun.}\ }\textbf {\bibinfo {volume} {224}},\
  \bibinfo {pages} {405} (\bibinfo {year} {2018})}\BibitemShut {NoStop}%
\bibitem [{\citenamefont {\"Oz\ifmmode~\mbox{\c{c}}\else \c{c}\fi{}elik}\ \emph
  {et~al.}(2015)\citenamefont {\"Oz\ifmmode~\mbox{\c{c}}\else \c{c}\fi{}elik},
  \citenamefont {Akt\"urk}, \citenamefont {Durgun},\ and\ \citenamefont
  {Ciraci}}]{hex-nitrogene}%
  \BibitemOpen
  \bibfield  {author} {\bibinfo {author} {\bibfnamefont {V.~O.}\ \bibnamefont
  {\"Oz\ifmmode~\mbox{\c{c}}\else \c{c}\fi{}elik}}, \bibinfo {author}
  {\bibfnamefont {O.~U.}\ \bibnamefont {Akt\"urk}}, \bibinfo {author}
  {\bibfnamefont {E.}~\bibnamefont {Durgun}}, \ and\ \bibinfo {author}
  {\bibfnamefont {S.}~\bibnamefont {Ciraci}},\ }\href {\doibase
  10.1103/PhysRevB.92.125420} {\bibfield  {journal} {\bibinfo  {journal} {Phys.
  Rev. B}\ }\textbf {\bibinfo {volume} {92}},\ \bibinfo {pages} {125420}
  (\bibinfo {year} {2015})}\BibitemShut {NoStop}%
\bibitem [{\citenamefont {Zhang}\ \emph {et~al.}(2019)\citenamefont {Zhang},
  \citenamefont {Kang}, \citenamefont {Huang}, \citenamefont {Jiang},
  \citenamefont {Ni}, \citenamefont {Kang}, \citenamefont {Zhang},
  \citenamefont {Xu}, \citenamefont {Liu},\ and\ \citenamefont
  {Liu}}]{dklPhysRevB.99.100404}%
  \BibitemOpen
  \bibfield  {author} {\bibinfo {author} {\bibfnamefont {S.}~\bibnamefont
  {Zhang}}, \bibinfo {author} {\bibfnamefont {M.}~\bibnamefont {Kang}},
  \bibinfo {author} {\bibfnamefont {H.}~\bibnamefont {Huang}}, \bibinfo
  {author} {\bibfnamefont {W.}~\bibnamefont {Jiang}}, \bibinfo {author}
  {\bibfnamefont {X.}~\bibnamefont {Ni}}, \bibinfo {author} {\bibfnamefont
  {L.}~\bibnamefont {Kang}}, \bibinfo {author} {\bibfnamefont {S.}~\bibnamefont
  {Zhang}}, \bibinfo {author} {\bibfnamefont {H.}~\bibnamefont {Xu}}, \bibinfo
  {author} {\bibfnamefont {Z.}~\bibnamefont {Liu}}, \ and\ \bibinfo {author}
  {\bibfnamefont {F.}~\bibnamefont {Liu}},\ }\href {\doibase
  10.1103/PhysRevB.99.100404} {\bibfield  {journal} {\bibinfo  {journal} {Phys.
  Rev. B}\ }\textbf {\bibinfo {volume} {99}},\ \bibinfo {pages} {100404}
  (\bibinfo {year} {2019})}\BibitemShut {NoStop}%
\bibitem [{\citenamefont {Cai}\ \emph {et~al.}(2023)\citenamefont {Cai},
  \citenamefont {Wang}, \citenamefont {Wang}, \citenamefont {Hao},
  \citenamefont {Liu}, \citenamefont {Zhou}, \citenamefont {Sui}, \citenamefont
  {Jiang}, \citenamefont {Xu}, \citenamefont {Ge}, \citenamefont {Ma},
  \citenamefont {Zhang}, \citenamefont {Shen}, \citenamefont {Yang},
  \citenamefont {Jiang}, \citenamefont {Liu}, \citenamefont {Ye}, \citenamefont
  {Shen}, \citenamefont {Liu}, \citenamefont {Cui}, \citenamefont {Wang},
  \citenamefont {Liu}, \citenamefont {Lin}, \citenamefont {Huang},
  \citenamefont {Wu}, \citenamefont {Zhuang}, \citenamefont {He}, \citenamefont
  {Zhang}, \citenamefont {Mei},\ and\ \citenamefont
  {Chen}}]{DKLaelm.202300212}%
  \BibitemOpen
  \bibfield  {author} {\bibinfo {author} {\bibfnamefont {Y.}~\bibnamefont
  {Cai}}, \bibinfo {author} {\bibfnamefont {J.}~\bibnamefont {Wang}}, \bibinfo
  {author} {\bibfnamefont {Y.}~\bibnamefont {Wang}}, \bibinfo {author}
  {\bibfnamefont {Z.}~\bibnamefont {Hao}}, \bibinfo {author} {\bibfnamefont
  {Y.}~\bibnamefont {Liu}}, \bibinfo {author} {\bibfnamefont {L.}~\bibnamefont
  {Zhou}}, \bibinfo {author} {\bibfnamefont {X.}~\bibnamefont {Sui}}, \bibinfo
  {author} {\bibfnamefont {Z.}~\bibnamefont {Jiang}}, \bibinfo {author}
  {\bibfnamefont {S.}~\bibnamefont {Xu}}, \bibinfo {author} {\bibfnamefont
  {H.}~\bibnamefont {Ge}}, \bibinfo {author} {\bibfnamefont {X.-M.}\
  \bibnamefont {Ma}}, \bibinfo {author} {\bibfnamefont {C.}~\bibnamefont
  {Zhang}}, \bibinfo {author} {\bibfnamefont {Z.}~\bibnamefont {Shen}},
  \bibinfo {author} {\bibfnamefont {Y.}~\bibnamefont {Yang}}, \bibinfo {author}
  {\bibfnamefont {Q.}~\bibnamefont {Jiang}}, \bibinfo {author} {\bibfnamefont
  {Z.}~\bibnamefont {Liu}}, \bibinfo {author} {\bibfnamefont {M.}~\bibnamefont
  {Ye}}, \bibinfo {author} {\bibfnamefont {D.}~\bibnamefont {Shen}}, \bibinfo
  {author} {\bibfnamefont {Y.}~\bibnamefont {Liu}}, \bibinfo {author}
  {\bibfnamefont {S.}~\bibnamefont {Cui}}, \bibinfo {author} {\bibfnamefont
  {L.}~\bibnamefont {Wang}}, \bibinfo {author} {\bibfnamefont {C.}~\bibnamefont
  {Liu}}, \bibinfo {author} {\bibfnamefont {J.}~\bibnamefont {Lin}}, \bibinfo
  {author} {\bibfnamefont {B.}~\bibnamefont {Huang}}, \bibinfo {author}
  {\bibfnamefont {L.}~\bibnamefont {Wu}}, \bibinfo {author} {\bibfnamefont
  {J.}~\bibnamefont {Zhuang}}, \bibinfo {author} {\bibfnamefont
  {H.}~\bibnamefont {He}}, \bibinfo {author} {\bibfnamefont {W.}~\bibnamefont
  {Zhang}}, \bibinfo {author} {\bibfnamefont {J.-W.}\ \bibnamefont {Mei}}, \
  and\ \bibinfo {author} {\bibfnamefont {C.}~\bibnamefont {Chen}},\ }\href
  {\doibase https://doi.org/10.1002/aelm.202300212} {\bibfield  {journal}
  {\bibinfo  {journal} {Adv. Electron. Mater.}\ }\textbf {\bibinfo {volume}
  {9}},\ \bibinfo {pages} {2300212} (\bibinfo {year} {2023})}\BibitemShut
  {NoStop}%
\bibitem [{\citenamefont {Liu}\ \emph {et~al.}(2021)\citenamefont {Liu},
  \citenamefont {Meng},\ and\ \citenamefont
  {Liu}}]{DKLPhysRevMaterials.5.084203}%
  \BibitemOpen
  \bibfield  {author} {\bibinfo {author} {\bibfnamefont {H.}~\bibnamefont
  {Liu}}, \bibinfo {author} {\bibfnamefont {S.}~\bibnamefont {Meng}}, \ and\
  \bibinfo {author} {\bibfnamefont {F.}~\bibnamefont {Liu}},\ }\href {\doibase
  10.1103/PhysRevMaterials.5.084203} {\bibfield  {journal} {\bibinfo  {journal}
  {Phys. Rev. Mater.}\ }\textbf {\bibinfo {volume} {5}},\ \bibinfo {pages}
  {084203} (\bibinfo {year} {2021})}\BibitemShut {NoStop}%
\bibitem [{\citenamefont {Jiang}\ \emph {et~al.}(2021)\citenamefont {Jiang},
  \citenamefont {Ni},\ and\ \citenamefont {Liu}}]{organicgrid}%
  \BibitemOpen
  \bibfield  {author} {\bibinfo {author} {\bibfnamefont {W.}~\bibnamefont
  {Jiang}}, \bibinfo {author} {\bibfnamefont {X.}~\bibnamefont {Ni}}, \ and\
  \bibinfo {author} {\bibfnamefont {F.}~\bibnamefont {Liu}},\ }\href {\doibase
  10.1021/acs.accounts.0c00652} {\bibfield  {journal} {\bibinfo  {journal}
  {Accounts Chem. Res.}\ }\textbf {\bibinfo {volume} {54}},\ \bibinfo {pages}
  {416} (\bibinfo {year} {2021})}\BibitemShut {NoStop}%
\bibitem [{\citenamefont {Chen}\ \emph {et~al.}(2018)\citenamefont {Chen},
  \citenamefont {Xu}, \citenamefont {Xie}, \citenamefont {Zhong}, \citenamefont
  {Wu},\ and\ \citenamefont {Zhang}}]{dklPhysRevB.98.035135}%
  \BibitemOpen
  \bibfield  {author} {\bibinfo {author} {\bibfnamefont {Y.}~\bibnamefont
  {Chen}}, \bibinfo {author} {\bibfnamefont {S.}~\bibnamefont {Xu}}, \bibinfo
  {author} {\bibfnamefont {Y.}~\bibnamefont {Xie}}, \bibinfo {author}
  {\bibfnamefont {C.}~\bibnamefont {Zhong}}, \bibinfo {author} {\bibfnamefont
  {C.}~\bibnamefont {Wu}}, \ and\ \bibinfo {author} {\bibfnamefont {S.~B.}\
  \bibnamefont {Zhang}},\ }\href {\doibase 10.1103/PhysRevB.98.035135}
  {\bibfield  {journal} {\bibinfo  {journal} {Phys. Rev. B}\ }\textbf {\bibinfo
  {volume} {98}},\ \bibinfo {pages} {035135} (\bibinfo {year}
  {2018})}\BibitemShut {NoStop}%
\bibitem [{\citenamefont {Zhu}\ \emph {et~al.}(2020)\citenamefont {Zhu},
  \citenamefont {He}, \citenamefont {Zhao},\ and\ \citenamefont
  {Fu}}]{bftC9TC06132K}%
  \BibitemOpen
  \bibfield  {author} {\bibinfo {author} {\bibfnamefont {J.}~\bibnamefont
  {Zhu}}, \bibinfo {author} {\bibfnamefont {C.}~\bibnamefont {He}}, \bibinfo
  {author} {\bibfnamefont {Y.-H.}\ \bibnamefont {Zhao}}, \ and\ \bibinfo
  {author} {\bibfnamefont {B.}~\bibnamefont {Fu}},\ }\href {\doibase
  10.1039/C9TC06132K} {\bibfield  {journal} {\bibinfo  {journal} {J. Mater.
  Chem. C}\ }\textbf {\bibinfo {volume} {8}},\ \bibinfo {pages} {2732}
  (\bibinfo {year} {2020})}\BibitemShut {NoStop}%
\bibitem [{\citenamefont {Hieu}\ \emph {et~al.}(2022)\citenamefont {Hieu},
  \citenamefont {Phuc}, \citenamefont {Kartamyshev},\ and\ \citenamefont
  {Vu}}]{mechanicalstability}%
  \BibitemOpen
  \bibfield  {author} {\bibinfo {author} {\bibfnamefont {N.~N.}\ \bibnamefont
  {Hieu}}, \bibinfo {author} {\bibfnamefont {H.~V.}\ \bibnamefont {Phuc}},
  \bibinfo {author} {\bibfnamefont {A.~I.}\ \bibnamefont {Kartamyshev}}, \ and\
  \bibinfo {author} {\bibfnamefont {T.~V.}\ \bibnamefont {Vu}},\ }\href
  {\doibase 10.1103/PhysRevB.105.075402} {\bibfield  {journal} {\bibinfo
  {journal} {Phys. Rev. B}\ }\textbf {\bibinfo {volume} {105}},\ \bibinfo
  {pages} {075402} (\bibinfo {year} {2022})}\BibitemShut {NoStop}%
\bibitem [{\citenamefont {Sethi}\ \emph {et~al.}(2023)\citenamefont {Sethi},
  \citenamefont {Cuma},\ and\ \citenamefont
  {Liu}}]{liufengPhysRevLett.130.186401}%
  \BibitemOpen
  \bibfield  {author} {\bibinfo {author} {\bibfnamefont {G.}~\bibnamefont
  {Sethi}}, \bibinfo {author} {\bibfnamefont {M.}~\bibnamefont {Cuma}}, \ and\
  \bibinfo {author} {\bibfnamefont {F.}~\bibnamefont {Liu}},\ }\href {\doibase
  10.1103/PhysRevLett.130.186401} {\bibfield  {journal} {\bibinfo  {journal}
  {Phys. Rev. Lett.}\ }\textbf {\bibinfo {volume} {130}},\ \bibinfo {pages}
  {186401} (\bibinfo {year} {2023})}\BibitemShut {NoStop}%
\bibitem [{\citenamefont {Li}\ \emph {et~al.}(2013)\citenamefont {Li},
  \citenamefont {Zhang}, \citenamefont {Wang}, \citenamefont {Yang},
  \citenamefont {Li}, \citenamefont {Zhu}, \citenamefont {Zhou}, \citenamefont
  {Han},\ and\ \citenamefont {Li}}]{li2013spatial}%
  \BibitemOpen
  \bibfield  {author} {\bibinfo {author} {\bibfnamefont {R.}~\bibnamefont
  {Li}}, \bibinfo {author} {\bibfnamefont {F.}~\bibnamefont {Zhang}}, \bibinfo
  {author} {\bibfnamefont {D.}~\bibnamefont {Wang}}, \bibinfo {author}
  {\bibfnamefont {J.}~\bibnamefont {Yang}}, \bibinfo {author} {\bibfnamefont
  {M.}~\bibnamefont {Li}}, \bibinfo {author} {\bibfnamefont {J.}~\bibnamefont
  {Zhu}}, \bibinfo {author} {\bibfnamefont {X.}~\bibnamefont {Zhou}}, \bibinfo
  {author} {\bibfnamefont {H.}~\bibnamefont {Han}}, \ and\ \bibinfo {author}
  {\bibfnamefont {C.}~\bibnamefont {Li}},\ }\href {\doibase 10.1038/ncomms2401}
  {\bibfield  {journal} {\bibinfo  {journal} {Nat. Commun.}\ }\textbf {\bibinfo
  {volume} {4}},\ \bibinfo {pages} {1432} (\bibinfo {year} {2013})}\BibitemShut
  {NoStop}%
\bibitem [{\citenamefont {Gajdo\ifmmode~\check{s}\else \v{s}\fi{}}\ \emph
  {et~al.}(2006{\natexlab{a}})\citenamefont {Gajdo\ifmmode~\check{s}\else
  \v{s}\fi{}}, \citenamefont {Hummer}, \citenamefont {Kresse}, \citenamefont
  {Furthm\"uller},\ and\ \citenamefont {Bechstedt}}]{linearresponsetheory}%
  \BibitemOpen
  \bibfield  {author} {\bibinfo {author} {\bibfnamefont {M.}~\bibnamefont
  {Gajdo\ifmmode~\check{s}\else \v{s}\fi{}}}, \bibinfo {author} {\bibfnamefont
  {K.}~\bibnamefont {Hummer}}, \bibinfo {author} {\bibfnamefont
  {G.}~\bibnamefont {Kresse}}, \bibinfo {author} {\bibfnamefont
  {J.}~\bibnamefont {Furthm\"uller}}, \ and\ \bibinfo {author} {\bibfnamefont
  {F.}~\bibnamefont {Bechstedt}},\ }\href {\doibase 10.1103/PhysRevB.73.045112}
  {\bibfield  {journal} {\bibinfo  {journal} {Phys. Rev. B}\ }\textbf {\bibinfo
  {volume} {73}},\ \bibinfo {pages} {045112} (\bibinfo {year}
  {2006}{\natexlab{a}})}\BibitemShut {NoStop}%
\bibitem [{\citenamefont {Gajdo\ifmmode~\check{s}\else \v{s}\fi{}}\ \emph
  {et~al.}(2006{\natexlab{b}})\citenamefont {Gajdo\ifmmode~\check{s}\else
  \v{s}\fi{}}, \citenamefont {Hummer}, \citenamefont {Kresse}, \citenamefont
  {Furthm\"uller},\ and\ \citenamefont {Bechstedt}}]{PhysRevB.73.045112}%
  \BibitemOpen
  \bibfield  {author} {\bibinfo {author} {\bibfnamefont {M.}~\bibnamefont
  {Gajdo\ifmmode~\check{s}\else \v{s}\fi{}}}, \bibinfo {author} {\bibfnamefont
  {K.}~\bibnamefont {Hummer}}, \bibinfo {author} {\bibfnamefont
  {G.}~\bibnamefont {Kresse}}, \bibinfo {author} {\bibfnamefont
  {J.}~\bibnamefont {Furthm\"uller}}, \ and\ \bibinfo {author} {\bibfnamefont
  {F.}~\bibnamefont {Bechstedt}},\ }\href {\doibase 10.1103/PhysRevB.73.045112}
  {\bibfield  {journal} {\bibinfo  {journal} {Phys. Rev. B}\ }\textbf {\bibinfo
  {volume} {73}},\ \bibinfo {pages} {045112} (\bibinfo {year}
  {2006}{\natexlab{b}})}\BibitemShut {NoStop}%
\bibitem [{\citenamefont {Matthes}\ \emph {et~al.}(2014)\citenamefont
  {Matthes}, \citenamefont {Pulci},\ and\ \citenamefont
  {Bechstedt}}]{matthes2014optical}%
  \BibitemOpen
  \bibfield  {author} {\bibinfo {author} {\bibfnamefont {L.}~\bibnamefont
  {Matthes}}, \bibinfo {author} {\bibfnamefont {O.}~\bibnamefont {Pulci}}, \
  and\ \bibinfo {author} {\bibfnamefont {F.}~\bibnamefont {Bechstedt}},\ }\href
  {\doibase 10.1088/1367-2630/16/10/105007} {\bibfield  {journal} {\bibinfo
  {journal} {New J. Phys.}\ }\textbf {\bibinfo {volume} {16}},\ \bibinfo
  {pages} {105007} (\bibinfo {year} {2014})}\BibitemShut {NoStop}%
\bibitem [{\citenamefont {Matthes}\ \emph {et~al.}(2016)\citenamefont
  {Matthes}, \citenamefont {Pulci},\ and\ \citenamefont
  {Bechstedt}}]{PhysRevB.94.205408}%
  \BibitemOpen
  \bibfield  {author} {\bibinfo {author} {\bibfnamefont {L.}~\bibnamefont
  {Matthes}}, \bibinfo {author} {\bibfnamefont {O.}~\bibnamefont {Pulci}}, \
  and\ \bibinfo {author} {\bibfnamefont {F.}~\bibnamefont {Bechstedt}},\ }\href
  {\doibase 10.1103/PhysRevB.94.205408} {\bibfield  {journal} {\bibinfo
  {journal} {Phys. Rev. B}\ }\textbf {\bibinfo {volume} {94}},\ \bibinfo
  {pages} {205408} (\bibinfo {year} {2016})}\BibitemShut {NoStop}%
\bibitem [{\citenamefont {Zhang}\ \emph {et~al.}(2015)\citenamefont {Zhang},
  \citenamefont {Lee}, \citenamefont {Wang},\ and\ \citenamefont
  {Yao}}]{oct-nitrogene}%
  \BibitemOpen
  \bibfield  {author} {\bibinfo {author} {\bibfnamefont {Y.}~\bibnamefont
  {Zhang}}, \bibinfo {author} {\bibfnamefont {J.}~\bibnamefont {Lee}}, \bibinfo
  {author} {\bibfnamefont {W.-L.}\ \bibnamefont {Wang}}, \ and\ \bibinfo
  {author} {\bibfnamefont {D.-X.}\ \bibnamefont {Yao}},\ }\href {\doibase
  10.1016/j.commatsci.2015.08.008} {\bibfield  {journal} {\bibinfo  {journal}
  {Comp. Mater. Sci.}\ }\textbf {\bibinfo {volume} {110}},\ \bibinfo {pages}
  {109} (\bibinfo {year} {2015})}\BibitemShut {NoStop}%
\bibitem [{\citenamefont {Bondarchuk}\ and\ \citenamefont
  {Minaev}(2017)}]{ZS-nitrogene}%
  \BibitemOpen
  \bibfield  {author} {\bibinfo {author} {\bibfnamefont {S.~V.}\ \bibnamefont
  {Bondarchuk}}\ and\ \bibinfo {author} {\bibfnamefont {B.~F.}\ \bibnamefont
  {Minaev}},\ }\href {\doibase 10.1016/j.commatsci.2017.03.007} {\bibfield
  {journal} {\bibinfo  {journal} {Comp. Mater. Sci.}\ }\textbf {\bibinfo
  {volume} {133}},\ \bibinfo {pages} {122} (\bibinfo {year}
  {2017})}\BibitemShut {NoStop}%
\bibitem [{\citenamefont {Cao}\ \emph {et~al.}(2015)\citenamefont {Cao},
  \citenamefont {Li},\ and\ \citenamefont {Louie}}]{Stonercriterion}%
  \BibitemOpen
  \bibfield  {author} {\bibinfo {author} {\bibfnamefont {T.}~\bibnamefont
  {Cao}}, \bibinfo {author} {\bibfnamefont {Z.}~\bibnamefont {Li}}, \ and\
  \bibinfo {author} {\bibfnamefont {S.~G.}\ \bibnamefont {Louie}},\ }\href
  {\doibase 10.1103/PhysRevLett.114.236602} {\bibfield  {journal} {\bibinfo
  {journal} {Phys. Rev. Lett.}\ }\textbf {\bibinfo {volume} {114}},\ \bibinfo
  {pages} {236602} (\bibinfo {year} {2015})}\BibitemShut {NoStop}%
\bibitem [{\citenamefont {Hortamani}\ \emph {et~al.}(2008)\citenamefont
  {Hortamani}, \citenamefont {Sandratskii}, \citenamefont {Kratzer},
  \citenamefont {Mertig},\ and\ \citenamefont
  {Scheffler}}]{stonerPhysRevB.78.104402}%
  \BibitemOpen
  \bibfield  {author} {\bibinfo {author} {\bibfnamefont {M.}~\bibnamefont
  {Hortamani}}, \bibinfo {author} {\bibfnamefont {L.}~\bibnamefont
  {Sandratskii}}, \bibinfo {author} {\bibfnamefont {P.}~\bibnamefont
  {Kratzer}}, \bibinfo {author} {\bibfnamefont {I.}~\bibnamefont {Mertig}}, \
  and\ \bibinfo {author} {\bibfnamefont {M.}~\bibnamefont {Scheffler}},\ }\href
  {\doibase 10.1103/PhysRevB.78.104402} {\bibfield  {journal} {\bibinfo
  {journal} {Phys. Rev. B}\ }\textbf {\bibinfo {volume} {78}},\ \bibinfo
  {pages} {104402} (\bibinfo {year} {2008})}\BibitemShut {NoStop}%
\bibitem [{\citenamefont {Benalcazar}\ \emph
  {et~al.}(2017{\natexlab{a}})\citenamefont {Benalcazar}, \citenamefont
  {Bernevig},\ and\ \citenamefont {Hughes}}]{SOTIWladimir2017}%
  \BibitemOpen
  \bibfield  {author} {\bibinfo {author} {\bibfnamefont {W.~A.}\ \bibnamefont
  {Benalcazar}}, \bibinfo {author} {\bibfnamefont {B.~A.}\ \bibnamefont
  {Bernevig}}, \ and\ \bibinfo {author} {\bibfnamefont {T.~L.}\ \bibnamefont
  {Hughes}},\ }\href {\doibase 10.1126/science.aah6442} {\bibfield  {journal}
  {\bibinfo  {journal} {Science}\ }\textbf {\bibinfo {volume} {357}},\ \bibinfo
  {pages} {61} (\bibinfo {year} {2017}{\natexlab{a}})}\BibitemShut {NoStop}%
\bibitem [{\citenamefont {Benalcazar}\ \emph
  {et~al.}(2017{\natexlab{b}})\citenamefont {Benalcazar}, \citenamefont
  {Bernevig},\ and\ \citenamefont {Hughes}}]{PhysRevB.96.245115}%
  \BibitemOpen
  \bibfield  {author} {\bibinfo {author} {\bibfnamefont {W.~A.}\ \bibnamefont
  {Benalcazar}}, \bibinfo {author} {\bibfnamefont {B.~A.}\ \bibnamefont
  {Bernevig}}, \ and\ \bibinfo {author} {\bibfnamefont {T.~L.}\ \bibnamefont
  {Hughes}},\ }\href {\doibase 10.1103/PhysRevB.96.245115} {\bibfield
  {journal} {\bibinfo  {journal} {Phys. Rev. B}\ }\textbf {\bibinfo {volume}
  {96}},\ \bibinfo {pages} {245115} (\bibinfo {year}
  {2017}{\natexlab{b}})}\BibitemShut {NoStop}%
\bibitem [{\citenamefont {Sheng}\ \emph {et~al.}(2019)\citenamefont {Sheng},
  \citenamefont {Chen}, \citenamefont {Liu}, \citenamefont {Chen},
  \citenamefont {Yu}, \citenamefont {Zhao},\ and\ \citenamefont
  {Yang}}]{PhysRevLett.123.256402}%
  \BibitemOpen
  \bibfield  {author} {\bibinfo {author} {\bibfnamefont {X.-L.}\ \bibnamefont
  {Sheng}}, \bibinfo {author} {\bibfnamefont {C.}~\bibnamefont {Chen}},
  \bibinfo {author} {\bibfnamefont {H.}~\bibnamefont {Liu}}, \bibinfo {author}
  {\bibfnamefont {Z.}~\bibnamefont {Chen}}, \bibinfo {author} {\bibfnamefont
  {Z.-M.}\ \bibnamefont {Yu}}, \bibinfo {author} {\bibfnamefont {Y.~X.}\
  \bibnamefont {Zhao}}, \ and\ \bibinfo {author} {\bibfnamefont {S.~A.}\
  \bibnamefont {Yang}},\ }\href {\doibase 10.1103/PhysRevLett.123.256402}
  {\bibfield  {journal} {\bibinfo  {journal} {Phys. Rev. Lett.}\ }\textbf
  {\bibinfo {volume} {123}},\ \bibinfo {pages} {256402} (\bibinfo {year}
  {2019})}\BibitemShut {NoStop}%
\bibitem [{\citenamefont {Liu}\ \emph {et~al.}(2019)\citenamefont {Liu},
  \citenamefont {Zhao}, \citenamefont {Liu},\ and\ \citenamefont
  {Wang}}]{SOTIgraphyneNL}%
  \BibitemOpen
  \bibfield  {author} {\bibinfo {author} {\bibfnamefont {B.}~\bibnamefont
  {Liu}}, \bibinfo {author} {\bibfnamefont {G.}~\bibnamefont {Zhao}}, \bibinfo
  {author} {\bibfnamefont {Z.}~\bibnamefont {Liu}}, \ and\ \bibinfo {author}
  {\bibfnamefont {Z.~F.}\ \bibnamefont {Wang}},\ }\href {\doibase
  10.1021/acs.nanolett.9b02719} {\bibfield  {journal} {\bibinfo  {journal}
  {Nano Lett.}\ }\textbf {\bibinfo {volume} {19}},\ \bibinfo {pages} {6492}
  (\bibinfo {year} {2019})}\BibitemShut {NoStop}%
\bibitem [{\citenamefont {Qian}\ \emph {et~al.}(2021)\citenamefont {Qian},
  \citenamefont {Liu},\ and\ \citenamefont {Yao}}]{SOTIPhysRevB.104.245427}%
  \BibitemOpen
  \bibfield  {author} {\bibinfo {author} {\bibfnamefont {S.}~\bibnamefont
  {Qian}}, \bibinfo {author} {\bibfnamefont {C.-C.}\ \bibnamefont {Liu}}, \
  and\ \bibinfo {author} {\bibfnamefont {Y.}~\bibnamefont {Yao}},\ }\href
  {\doibase 10.1103/PhysRevB.104.245427} {\bibfield  {journal} {\bibinfo
  {journal} {Phys. Rev. B}\ }\textbf {\bibinfo {volume} {104}},\ \bibinfo
  {pages} {245427} (\bibinfo {year} {2021})}\BibitemShut {NoStop}%
\bibitem [{\citenamefont {Ahn}\ \emph {et~al.}(2019)\citenamefont {Ahn},
  \citenamefont {Park}, \citenamefont {Kim}, \citenamefont {Kim},\ and\
  \citenamefont {Yang}}]{Ahn2019}%
  \BibitemOpen
  \bibfield  {author} {\bibinfo {author} {\bibfnamefont {J.}~\bibnamefont
  {Ahn}}, \bibinfo {author} {\bibfnamefont {S.}~\bibnamefont {Park}}, \bibinfo
  {author} {\bibfnamefont {D.}~\bibnamefont {Kim}}, \bibinfo {author}
  {\bibfnamefont {Y.}~\bibnamefont {Kim}}, \ and\ \bibinfo {author}
  {\bibfnamefont {B.-J.}\ \bibnamefont {Yang}},\ }\href {\doibase
  10.1088/1674-1056/ab4d3b} {\bibfield  {journal} {\bibinfo  {journal} {Chin.
  Phys. B}\ }\textbf {\bibinfo {volume} {28}},\ \bibinfo {pages} {117101}
  (\bibinfo {year} {2019})}\BibitemShut {NoStop}%
\bibitem [{\citenamefont {Zhao}\ and\ \citenamefont
  {Lu}(2017)}]{PhysRevLett.118.056401}%
  \BibitemOpen
  \bibfield  {author} {\bibinfo {author} {\bibfnamefont {Y.~X.}\ \bibnamefont
  {Zhao}}\ and\ \bibinfo {author} {\bibfnamefont {Y.}~\bibnamefont {Lu}},\
  }\href {\doibase 10.1103/PhysRevLett.118.056401} {\bibfield  {journal}
  {\bibinfo  {journal} {Phys. Rev. Lett.}\ }\textbf {\bibinfo {volume} {118}},\
  \bibinfo {pages} {056401} (\bibinfo {year} {2017})}\BibitemShut {NoStop}%
\bibitem [{\citenamefont {Wieder}\ and\ \citenamefont
  {Bernevig}(2018)}]{wieder2018axion}%
  \BibitemOpen
  \bibfield  {author} {\bibinfo {author} {\bibfnamefont {B.~J.}\ \bibnamefont
  {Wieder}}\ and\ \bibinfo {author} {\bibfnamefont {B.~A.}\ \bibnamefont
  {Bernevig}},\ }\href {\doibase 10.48550/arXiv.1810.02373} {\bibfield
  {journal} {\bibinfo  {journal} {arXiv.1810.02373}\ } (\bibinfo {year}
  {2018}),\ 10.48550/arXiv.1810.02373}\BibitemShut {NoStop}%
\bibitem [{\citenamefont {Hu}\ \emph {et~al.}(2022)\citenamefont {Hu},
  \citenamefont {Zhang}, \citenamefont {Mu},\ and\ \citenamefont
  {Wang}}]{SOTIwangzf2022}%
  \BibitemOpen
  \bibfield  {author} {\bibinfo {author} {\bibfnamefont {T.}~\bibnamefont
  {Hu}}, \bibinfo {author} {\bibfnamefont {T.}~\bibnamefont {Zhang}}, \bibinfo
  {author} {\bibfnamefont {H.}~\bibnamefont {Mu}}, \ and\ \bibinfo {author}
  {\bibfnamefont {Z.}~\bibnamefont {Wang}},\ }\href {\doibase
  10.1021/acs.jpclett.2c02683} {\bibfield  {journal} {\bibinfo  {journal} {J.
  Phys. Chem. Lett.}\ }\textbf {\bibinfo {volume} {13}},\ \bibinfo {pages}
  {10905} (\bibinfo {year} {2022})}\BibitemShut {NoStop}%
\bibitem [{\citenamefont {Jiang}\ and\ \citenamefont
  {Park}(2014)}]{BPPoissonratio}%
  \BibitemOpen
  \bibfield  {author} {\bibinfo {author} {\bibfnamefont {J.-W.}\ \bibnamefont
  {Jiang}}\ and\ \bibinfo {author} {\bibfnamefont {H.~S.}\ \bibnamefont
  {Park}},\ }\href {\doibase 10.1038/ncomms5727} {\bibfield  {journal}
  {\bibinfo  {journal} {Nat. Commun.}\ }\textbf {\bibinfo {volume} {5}},\
  \bibinfo {pages} {4727} (\bibinfo {year} {2014})}\BibitemShut {NoStop}%
\end{thebibliography}%
\end{document}